\documentclass[]{aa} 
\usepackage{graphicx}
\usepackage{amsmath, amssymb}
\begin{document}
%
\def\hi {H\,{\sc i}}
\def\oiii {O{\sc iii}}
\def\hdueo {H$_2$O}
\def\fri {FR\,I}
\def\frii {FR\,II}
\def\txs {TXS\,2226{\tt -}184}
\def\pks {PKS\,2322{\tt -}123}
\def\radm {rad m$^{-2}$}
\def\ab {$\sim$}
\def\etal {{\sl et~al.\ }}
\def\dg{$^{\circ}$}
\def\kms{km\,s$^{-1}$}
\def\solmass {\hbox{M$_{\odot}$}}
\def\solum {\hbox{L$_{\odot}$}} 
\def\ffam {\hbox{$\,.\!\!^{\prime}$}}
\def\ffas {\hbox{$\,.\!\!^{\prime\prime}$}}
\def\ffM {\hbox{$\,.\!\!^{\rm M}$}}
\def\ffm {\hbox{$\,.\!\!^{\rm m}$}}
\def\ffs {\hbox{$\,.\!\!^{\rm s}$}}
\title{New \hdueo \,masers in Seyfert and FIR bright galaxies. III.\\
The Southern Sample.}

\author{G.\ Surcis \inst{1,2,3}
  \and 
   A.\ Tarchi \inst{1,4}
  \and
   C.\ Henkel \inst{3}
  \and
   J.\ Ott \thanks{JO is a Jansky Fellow of the National Radio Astronomy Observatory.}\inst{,5,6}
  \and
   J.\ Lovell \inst{7}
  \and
   P.\ Castangia \inst{1}}

\offprints{G. Surcis}

\institute{INAF-Osservatorio Astronomico di Cagliari, Loc. Poggio dei Pini, Strada 54, 09012 Capoterra (CA), Italy
 \and
 Argelander-Institut f\"{u}r Astronomie der Universit\"{a}t Bonn, Auf dem H\"{u}gel 71, 53121 Bonn, Germany\\
 \email{gsurcis@astro.uni-bonn.de}
 \and
 Max-Planck Institut f\"{u}r Radioastronomie, Auf dem H\"{u}gel 69, 53121 Bonn, Germany
 \and
 INAF-Istituto di Radioastronomia, Via Gobetti 101, 40129 Bologna, Italy
\and
 National Radio Astronomy Observatory, P.O. Box O, 1003 Lopezville Road, Socorro, NM 87801-0387, USA
 \and 
 California Institute of Technology, 1200 E California Blvd, Caltech Astronomy 105-24, Pasadena, CA 91125, USA
 \and
 Department of Maths and Physics, University of Tasmania, Private Bag 21, Hobart, Tasmania 7001, Australia}

\date{Received ; accepted}
\abstract
{Recently, a relationship between the water maser detection rate and far infrared (FIR) flux densities has been established as a result of two 22\,GHz maser surveys in a complete sample of galaxies ($\rm{Dec>-30^{\circ}}$) with $\rm{100 \, \mu m}$ flux densities of $> 50$\,Jy and $> 30$\,Jy.}
{This survey has been extended to the southern galaxies in order to discover new maser sources and to investigate the galaxies hosting the maser spots with particular emphasis on their nuclear regions.} 
{A sample of 12 galaxies with $\rm{Dec<-30^{\circ}}$ and $\rm{S_{100 \, \mu m}>50 \, Jy}$ was observed with the 70-m telescope of the Canberra Deep Space Communication Complex (CDSCC) at Tidbinbilla (Australia) in a search for water maser emission. The average 3$\sigma$ noise level of the survey is 15 mJy for a $\rm{0.42 \,km\, s^{-1}}$ channel, corresponding to a detection threshold of $\rm{\sim 0.1\, \solum}$ for the isotropic maser luminosity at a distance of 25\,Mpc.}
{Two new detections are reported: a kilomaser with an isotropic luminosity $L_{\rm{H_{2}O}}\sim5\, \solum$ in NGC\,3620 and a maser with about twice this luminosity in the merger system NGC\,3256. The detections have been followed-up through continuum and spectral line interferometric observations with the Australia Telescope Compact Array (ATCA). In NGC\,3256, a fraction (about a third) of the maser emission arises from two hot spots associated with star formation activity, which are offset from the galactic nuclei of the system. The remaining emission may arise from weaker centers of maser activity distributed over the central $50''$. For NGC\,3620, the water maser is coincident with the nuclear region of the galaxy. Our continuum observations indicate that the nature of the nuclear emission is most likely linked to particularly intense star formation. Including the historical detection in NGC\,4945, the water maser detection rate in the southern sample is 15\% (3/20), consistent with the northern sample. The high rate of maser detections in the complete all-sky FIR sample (23\%, 15/65) confirms the existence of a link between overall FIR flux density and maser phenomena. A relation between \hdueo \, and OH masers in the FIR sample is also discussed.}
{}
\keywords{Galaxies: individual: NGC\,3620, NGC\,3256 -- 
     Galaxies: active -- 
     Galaxies: ISM -- 
     masers -- 
     Radio lines: ISM -- 
     Radio lines: galaxies}

\titlerunning{New \hdueo \,masers in Seyfert and FIR bright galaxies. III.}
\authorrunning{Surcis et al.}

\maketitle
\section{Introduction}\label{intro}
A crucial problem in the study of active galaxies is that of understanding the detailed  geometry, physics and evolution of the central engines and their environments. Another longstanding and unexplored problem is the measurement of proper motions and geometrical distances of galaxies. In both fields, observations of water masers at 22 GHz ($\rm{\lambda\sim1.3\,cm}$) are indispensable. The rest frequency corresponds to the strongest water maser transition, the $6_{16} \rightarrow 5_{23}$ line of ortho-\hdueo, which originates in dense ($n(\rm{H_{2}})\gtrsim\rm{10^7}$ $\rm{cm^{-3}}$) and warm ($T_{\rm{kin}}\gtrsim\rm{400 \,K}$) molecular gas (e.g., Henkel et al., \cite{henkel05a}).\\
\indent In literature, extragalactic \hdueo \, masers are commonly classified according to their isotropic luminosity: the ``megamasers'' with $L_{\rm{H_{2}O}}>10 \,\rm{L_{\odot}}$ and the ``kilomasers'' with $\rm{L_{H_{2}O}<10 \,L_{\odot}}$. Because of the beamed nature of the maser emission, this classification may not in every case carry a significant physical meaning. However, for the sake of consistency with previous studies we will also adopt the aforementioned classification throughout our paper. The ``megamasers'' are seen out to cosmological distances (Impellizzeri et al., \cite{imp08}), and, when studied in detail, are always found to be associated with Active Galactic Nuclei (AGN) (e.g. Lo \cite{lo05}). The ``kilomasers'' are instead mostly associated with star forming regions similar to those seen in the Galaxy. In the latter class of objects, however, a few sources (e.g. M\,51: Hagiwara et al. \cite{hagiwara01}; NGC\,4051: Hagiwara et al. \cite{hagiwara03}; NGC\,4151: Braatz et al. \cite{braatz04}) also show hints of an association with the nucleus of their host galaxies, unveiling the possible presence of a buried AGN of low luminosity (LLAGN) (Hagiwara et al. \cite{hagiwara01}) and hinting at the possibility that some kilomasers also belong to the family of AGN masers, only representing the weak tail of their distribution as postulated by Ho et al. (\cite{Ho87}).\\ 
\indent In the last two decades 22 GHz \hdueo \, maser surveys of various samples of galaxies were carried out (e.g., Henkel et al. \cite{henkel86};  Braatz et al. \cite{braatz96}, \cite{braatz97}; Greenhill et al. \cite{green03}; Hagiwara et al. \cite{hagiwara03}; Henkel et al. \cite{henkel05b} (HPT); Kondratko et al. \cite{kondra06}; Braatz \& Gugliucci \cite{braatz08}; Castangia et al. \cite{casta08} (CTH)). The survey of Braatz et al. (\cite{braatz97}) was composed of 354 active galaxies, including  a distance-limited sample of Seyferts and Low Ionization Nuclear Emission Regions (LINERs) with recessional velocities $\rm{<7000 \, km s^{-1}}$, and a magnitude-limited sample, also covering Seyferts and LINERs, with optical magnitudes $\rm{m_{B}\leq14.5}$. Their 10 newly detected sources made it possible to associate the \hdueo \, megamaser phenomenon with Seyfert 2 and LINER galaxies and not with Seyfert 1 or other types of galaxies. This indicates that obscuring gas around an active nucleus, coupled with long gain paths along a preferentially edge-on oriented circumnuclear disk plays an important role.\\
\indent Detecting more extragalactic water masers is important to tackle a number of important astrophysical problems. In particular, as previously mentioned, these sources can be used to determine proper motions in the Local Group (Brunthaler et al. \cite{bru05}, \cite{bru07}), pinpoint sites of active star formation (e.g. Tarchi et al. \cite{tarchi02}; Brunthaler et al. \cite{bru06}), obtain geometrical distances, map accretion disks surrounding AGN (Miyoshi et al. \cite{miy95}; Herrnstein et al. \cite{her99}; Reid et al. \cite{reid09}), and, ultimately, improve our knowledge on the equation of state of the dark energy by obtaining a high precision measurement of the Hubble constant (Braatz et al. \cite{braatz008}).\\
\indent Since we know that far-infrared (FIR) emission commonly arises from dust grains heated by newly formed stars, a sample of FIR bright galaxies is a suitable tool to detect extragalactic \hdueo \, masers, in particular those associated with star forming regions. A correlation between the presence of \hdueo \, masers and $\rm{100 \, \mu m}$ flux density ($\rm{S_{100 \, \mu m}>50 \, Jy}$) was indeed reported by HPT as a result of an Effelsberg 100-m survey of all galaxies with declination $\rm{Dec>-30^{\circ}}$, associated with IRAS point sources showing $\rm{S_{100 \, \mu m}>50 \, Jy}$ (Fullmer \& Lonsdale \cite{fullmer89}) (hereafter referred to as ``the northern sample''). They obtained a detection rate of 22 \% \,(10/45) and found a correlation between FIR flux density and the luminosity of extragalactic maser emission. This correlation has recently been confirmed by CTH with a complementary sample of 41 galaxies with $\rm{30 \, Jy<S_{100 \, \mu m}<50 \, Jy}$ and $\rm{Dec>-30^{\circ}}$, that yielded a detection rate of 5\% in agreement with the expected value derived by the correlation. This is corroborated by the recent particularly sensitive study of Darling et al. (\cite{darl08}), which expanded the sample of known \hdueo \, kilomasers associated with star formation and ascribed the paucity of detections for the weaker tail of the extragalactic water masers mostly to the lack of sufficient sensitivity of the past surveys.\\
\indent The high number of detections obtained by HPT and CTH have motivated us to complete the whole sample by including all southern galaxies (with declination $\rm{Dec<-30^{\circ}}$) associated with IRAS point sources characterized by $\rm{S_{100 \, \mu m}>50 \, Jy}$ (hereafter referred to as ``the southern sample'').
\section{Observations and data reduction}\label{obssect}
The list of sources of our sample, composed of 20 galaxies, was compiled using the IRAS Point Source Catalog (Fullmer \& Lonsdale \cite{fullmer89}) and is shown in Table \ref{southsample}. Eight of these sources were already part of previous surveys and in the table their names are given in italics. Names of the twelve remaining sources are boldfaced.
\begin {table}[h!]
\caption []{Southern Sample$^{a}$}
\begin{center}
\scriptsize
\begin{tabular}{l c c c c c}
\hline
\hline
\\
Source$^b$       & V-range            & rms         &$L_{\rm{H_{2}O}}^c$& Epoch      & Tel$^d$\\ 
                 &($\rm{km \ s^{-1}}$)& ($\rm{mJy}$)&  (\solum)         &            &      \\
\\
\hline
\\
\textbf{NGC 0055}& -592, 752          & 10          & $<0.001$          &25-Jan-2005 & T\\
\textbf{NGC 0134}&  891, 2236         & 5           & $<0.08$           &09-Sep-2007 & T\\
\textbf{NGC 0300}& -518, 827          & 9           & $<0.001$          &24-Apr-2005 & T\\
                 &                    &             &                   &30-Apr-2005 &\\
                 & -547, 797          & 5           & $<0.001$          &17-Sep-2007 &\\
                 &                    &             &                   &18-Sep-2007 &\\
\textit{NGC 1097}&                    & 15          & $<0.30$           &13-Aug-2003 & \\
\textbf{NGC 1313}& -262, 1083         & 4           & $<0.005$          &02-Jan-2005 & T\\
\textit{NGC 1365}&                    & 12          & $<0.39$           &09-Jun-2003 & \\
\textbf{NGC 1559}&  596, 1941         & 5           & $<0.05$           &09-Apr-2005 & T\\
                 &                    &             &                   &10-Apr-2005 &\\
\textit{NGC 1672}&                    & 19          & $<0.41$           &27-Jun-2004 & \\
\textbf{NGC 1792}& 478, 1823          & 5           & $<0.05$           &24-Apr-2005 & T\\
                 &                    &             &                   &30-Apr-2005 & \\
\textit{NGC 1808}&                    & 15          & $<0.18$           &05-May-2003 &\\
\textbf{NGC 2997}& 353, 1697          & 7           & $<0.05$           &17-Apr-2005 & T\\
                 &                    &             &                   &19-Apr-2005 &\\
                 &                    &             &                   &24-Apr-2005 &\\
\textbf{NGC 3256$^e$}& see Table 2    &             &                   &            & T\\
		 &		      &		    &			&	     & A\\
\textbf{NGC 3620$^e$}& see Table 2    &             &                   &            & T\\
		 &		      &		    &			&	     & A\\
\textbf{NGC 3621}& 35, 1380           & 5           & $<0.02$           &26-Mar-2005 & T\\
                 &                    &             &                   &02-Apr-2005 & \\
\textit{NGC 4945}&                    &             &  160              &20-Sep-1978 & \\
                 &                    &             &                   &22-Sep-1978 & \\
\textbf{NGC 5128$^f$}&-50, 1200      &   -         & -                 &10-Apr-2005 & T \\
                 &                    &             &                   &17-Apr-2005 &\\
                 &                    &             &                   &24-Apr-2005 & \\
                 &407, 668            & 2           &  $<0.02$          &29-Jun-2006 & A\\
\textit{NGC 6744}&                    & 15          & $<0.13$           &17-May-2002 & \\
\textit{NGC 7552}&                    & 13          & $<0.41$           &19-Jun-2004 &\\
\textit{NGC 7582}&                    & 13          & $<0.40$           &27-Jun-2004 &\\
\textbf{NGC 7793}& -445, 899          & 7           & $<0.002$          &26-Aug-2007 & T\\
                 &                    &             &                   &31-Aug-2005 &\\
\\
\hline
\end{tabular}
\end{center}
\scriptsize{$^{a}$ For coordinates, systemic velocities, 100 $\rm{\mu m}$ flux densities, and references see the Table containing the entire sample of FIR bright galaxies.\\
$^b$ Bold-faced source names refer to sources observed in this study; source names in italic refer to previously measured targets. For NGC\,4945, see Dos Santos \& L\`{e}pine (\cite{santos79}), Batchelor et al. (\cite{batche82}), Greenhill et al. (\cite{green97}). For the other sources in italics, see Kondrakto et al. (\cite{kondra06}). \\
$^c$ Upper limits for maser luminosities are derived from $L_{\rm{H_{2}O}}/[\solum]=0.023 \times S/[\rm{Jy}] \times \Delta \textit{v}/[\rm{km \ s^{-1}}] \times \textit{D}^{2}/[\rm{Mpc^{2}}]$, where \textit{S} is 3 times the rms, in column 4, and $\Delta v$ is the channel width, $\rm{0.5 \ km\  s^{-1}}$ and $\rm{1.7 \ km \ s^{-1}}$ in our Tidbinbilla and ATCA observations (NGC\,5128/Cen A only), respectively.\\
$^d$ Telescope: T = Tidbinbilla; A = ATCA.\\
$^e$ Newly detected maser.\\
$^f$ Observed with ATCA, because the 22 GHz continuum was too strong to obtain single-dish spectra with high baseline quality.}
\label{southsample}
\end{table}
\subsection{Tidbinbilla}
\indent The sources of our sub-sample were observed in the $6_{16} - 5_{23}$ transition of \hdueo\,(rest frequency: 22.23508\,GHz) with the 70m NASA Deep Space Network antenna located at Tidbinbilla\footnote{The Deep Space Network DSS-43 Tidbinbilla antenna is managed by the Jet Propulsion Laboratory, California Institute of Technology, under a contract with the National Aeronautics and Space Administration. Access for radio astronomy is provided through an arrangement between the Australian and US governments and is coordinated by the CSIRO Australia Telescope National Facility.}, Australia. Our observations were made in 2005, from January to April, and in 2007, from August to September. Some sources, like NGC\,3620, were observed in both periods. The full width at half maximum (FWHM) beamwidth was $\approx50''$ and the pointing accuracy was better than $20''$. In both periods, observations were made with the same correlator configuration consisting of two independently tuned left circular polarized (LCP) bands, each 64 MHz wide, and providing spectra with 2048 channels per band. Observations were made in a position switching mode. The integration time (including overheads) was typically 2.5 hours per galaxy which resulted in approximately 1 hour on source.\\
\indent Each IF covers a velocity range of $\rm{\sim 900 \,km \,s^{-1}}$ with a channel spacing of $\rm{0.42 \,km\, s^{-1}}$. The two IFs are centered at slightly different central frequencies, with a frequency separation of $\rm{\sim 50 \,MHz}$, leading to an overlap between the IFs of about $\rm{300 \,km\, s^{-1}}$. In this way we could sample a total velocity range of $\rm{\sim 1200 \,km\, s^{-1}}$. All data were reduced using standard procedures of the Australia Telescope National Facility (ATNF) Spectral Analysis Package version 2.2 (ASAP v2.2) which is based on AIPS++ software libraries (http://www.atnf.csiro.au/computing/software/asap/). In order to reduce the noise in the final spectrum, the scans of an individual undetected source were averaged, which implies a monitoring timescale of 30 days or less. In order to calibrate data, i.e. to convert them from Kelvin to Jansky, we used the standard equation provided by the ATNF,
\begin{equation}
S_{\nu}({\rm{Jy}})= T_{source} \cdot \chi_{c}=T_{source} \cdot \frac{C \cdot C_0 \cdot 0.7172}{G(el)},
\end{equation}
\noindent
where $C_0$ is the opacity correction and $G(el)$ is the gain curve at 22 GHz depending on the elevation. In our observations $\chi_{c}$ has a value of 1.75 for each period and the flux calibration uncertainty is estimated to be of order 20\%. More observational details are given in Table \ref{southsample}.\\
\indent The galaxies NGC\,3256 and NGC\,3620 were again observed in 2008, with the same correlator configuration. In March 2008 both IFs of NGC\,3620 were centered at the maser velocity (see Sect. 3.2) covering a velocity range of $\rm{\sim 900 \,km \,s^{-1}}$ with a channel spacing of $\rm{0.42 \,km\, s^{-1}}$; in April 2008, NGC\,3620 and NGC\,3256 were observed once more using the original correlator configuration covering a total $\rm{\sim 1200 \,km\, s^{-1}}$. Eight contiguous channels were averaged in the spectra presented in Sect\,3. 
\subsection{ATCA}
\begin {table*}[t]
\caption []{Line parameters of newly detected masers.}
\begin{center}
\scriptsize
\begin{tabular}{l c c c c c c c c c l}
\hline
\hline
\\
Source &\textit{D}$^{a}$&Epoch&Tel.&$\int S \,\rm{d} \textit{V}^{b}$&$V_{\rm{LSR,opt}}^{b}$ &$\Delta V_{1/2}^{b}$&\textit{V}-range&Channel&rms&$L_{\rm{H_{2}O}}^c$ \\ 
       &      &     &    &                &                &                  &        & Width  &   &  \\
       &(Mpc) &     &    & (mJy km s$^{-1}$)& (km s$^{-1}$)    & (km s$^{-1}$)      &(km s$^{-1}$)& (km s$^{-1}$)& (mJy)&(\solum) \\
\\
\hline
\\
NGC\,3256 & 37.4& Aug 2007      & Tid &315  & $2827\pm6$      &    $63\pm14$       & 2065, 3407& 0.5 & 3 & 10.9\\
          &     & Apr 2008      & ATCA&27.0 & $2822.1\pm0.9$  &    $7.7\pm2.3$     & 2611, 2878& 3.7 & 4 & 1.0\\
          &     &               &     &32.8 & $2840.2\pm2.4$  &    $15.6\pm6.3$    &           &     & 7 & 2.4\\
          &     & Apr 2008      & Tid &400  & $2797\pm2$      &    $40.0^{d}$      & 2067, 3409& 0.5 & 3 & 12.9\\ 
          &     &               &     &         &           &                  &           &     &   & \\
NGC\,3620 & 22.4& Feb-March 2005& Tid &150      & $1823^{d}$&    $10.0^{d}$    & 1000, 2346& 0.5 & 7& 4.7$^{e}$ \\
          &     &               &     &255.2    & $1833^{d}$&   $4.4\pm0.5$     &           &     &   & \\
          &     & Apr 2006      &ATCA & 71.3 & $1821.1\pm0.1$ & $2.5\pm0.2$    & 1777, 1854& 0.2 & 3 & 2.1$^{f}$\\
          &     &               &     & 12.7 & $1825.1\pm0.1$ &   $1.4\pm0.4$  &           &     &   & \\
          &     &               &     & 97.2 & $1827.4\pm0.8$ &  $12.3\pm1.4$  &           &     &   & \\
          &     & Aug 2007      & Tid & 376  & $1826.1\pm0.8$ & $20.0^{d}$     & 952, 2298 & 0.5 & 3 & 4.3\\
          &     & Mar 2008      & Tid & 103.7&  $1820^{d}$  &  $6.1\pm1.7$    & 1385, 2245 & 0.5 & 5 & 3.2$^{e}$ \\
          &     &               &     & 124.7&  $1828^{d}$  &  $4.3\pm0.8$    &            &     &   &  \\
          &     & Apr 2008      & Tid & 114.4& $1817.5\pm3.4$ & $10.4\pm6.8$  &  994, 2339 & 0.5 & 5 & 2.5$^{e}$\\
          &     &               &     & 98.8 & $1824.7\pm0.8$ & $3.8\pm2.6$   &            &     &   & \\
\\
\hline 
\end{tabular}
\end{center}
\scriptsize{$^a$ Distances have been estimated by using velocities taken from the NASA/IPAC Extragalactic Database (NED) and adopting a Hubble costant of $H_{\rm{0}}= \rm{75 \, km\, s^{-1}\, Mpc^{-1}}$.\\
$^b$ Integrated flux densities, centre velocities, and full width to half maximum (FWHM) linewidths of the observed features are obtained from Gaussian fits.\\
$^c$ Isotropic luminosities are derived from $L_{\rm{H_{2}O}}/[\solum]=0.023 \times \int S \, \rm{d}\textit{V}/[\rm{Jy \ km \ s^{-1}}] \times \textit{D}^{2}/[\rm{Mpc^{2}}]$.\\
$^{d}$ Values obtained by forcing the fit.\\
$^e$ The water maser emission is composed of two features.\\
$^{f}$ The water maser emission is composed of three features.\\ }
\label{param}
\end{table*}

\begin {table*}[t]
\caption []{Observing parameters for the interferometric continuum maps of NGC\,3256 and NGC\,3620.} 
\begin{center}
\scriptsize
\begin{tabular}{l l c c c c c c  c}
\hline
\hline
\\
Source&Band    &$\nu$   &Observation &Telescope &Weighting &Restoring Beam &Position         & rms \\ 
     &  &(GHz) &date        &          &          & FWHM ($''$)   &Angle ($^{\circ}$)        & (mJy/beam)\\
\\
\hline
\\
NGC\,3256 &K       & 19.5 & 12-Apr-2008& ATCA     & Natural  & $1.19 \times 0.59$& 4.0     &  0.1\\
          &        &      & 13-Apr-2008&          &          &                   &         & \\
          &K$^{a}$ & 22.1 & 12-Apr-2008& ATCA     & Natural  & $1.16 \times 0.53$& 6.9     &  0.2 \\
          &        &      & 13-Apr-2008&          &          &                   &         &   \\
          &        &      &            &          &          &                   &         &\\
NGC\,3620 &C       & 4.8  & 14-Apr-2006& ATCA     & Natural  & $2.98 \times 2.51$& -4.6    &  0.4 \\
          &X       & 8.6  & 14-Apr-2006& ATCA     & Natural  & $1.92 \times 1.50$& -6.2    &  0.2 \\
          &K       & 19.5 & 15-Apr-2006& ATCA     & Natural  & $0.91 \times 0.61$& 12.1    &  0.2\\
          &        &      & 16-Apr-2006&          &          &                   &         &  \\
          &K$^{a}$ & 22.1 & 15-Apr-2006& ATCA     & Natural  & $0.85 \times 0.56$& 7.8     &  0.2 \\
          &        &      & 16-Apr-2006&          &          &                   &         &   \\
\\
\hline
\end{tabular}
\end{center}
\scriptsize{$^{a}$ Obtained by the line-free channels of our K-band data at 22 GHz.}\\
\label{conttab}
\end{table*}

NGC\,3620 was observed in continuum and spectral line mode with the Australia Telescope Compact Array (ATCA)\footnote{The Australia Telescope Compact Array is part of the Australia Telescope which is funded by the Commonwealth of Australia for operation as a National Facility managed by CSIRO}, an array of six 22-m antennas, in its most extended 6 km array configuration during average-good weather conditions in April 2006.\\
\indent Continuum data were obtained simultaneously at C and X-band, i.e. at 4799.907 and 8639.948 MHz, respectively, with a standard bandwidth of 128 MHz for a total integration time of 12 hours, including overheads. PKS1934-638 ($\rm{5.83\, Jy}$ at 4.8 GHz and $ \rm{2.84 \, Jy}$ at 8.6 GHz) and PKS1057-797 ($\rm{2.28 \, Jy}$ and $\rm{2.20 \,Jy}$, respectively) were used as amplitude and phase calibrators. Line and continuum data were reduced using standard procedures of the Multichannel Image Reconstruction, Image Analysis and Display (MIRIAD) data reduction package. This includes natural weighting and deconvolution by the CLEAN algorithm (H\"{o}gbom \cite{hogbom74}). \\
\indent NGC\,3620 was also observed in spectral line mode at 22.100 GHz. We have used the FULL\_8\_512\_128 correlator configuration with one IF (Frequency 1) centered on the maser line with a bandwidth of 8 MHz ($\rm{\sim 108\, km\, s^{-1}}$) and 512 spectral channels (channel spacing $\sim 16$\,kHz $\approx0.2$\,km\,s$^{-1}$). Simultaneously, the second IF (Frequency 2) was set to observe in continuum mode with a 128 MHz bandwidth at a central frequency of 19.5 GHz. The calibrators PKS1934-638 ($\rm{0.96\, Jy}$) and PKS1057-797 ($\rm{1.73 \, Jy}$) were observed for flux and phase calibration, respectively. The source PKS1921-293 ($\rm{12.67\, Jy}$) was used to calibrate the bandpass. Standard procedures implemented in MIRIAD were used to reduce the spectral data. We have subtracted the continuum by using the MIRIAD task UVLIN, which separates line and continuum in a spectral visibility data set, obtaining both a continuum-free cube and a continuum data set at 22.1 GHz.\\
\indent Toward the strong continuum source NGC\,5128 (Cen\,A) standing waves corrupted the single dish spectra during our survey at Tidbinbilla. In June 2006, NGC\,5128 (Cen \,A) we therefore observed NGC\,5128 with the ATCA 1.5 km array configuration in spectral line mode. The data were obtained at K-band (22 GHz) using the FULL\_32\_256 correlator configuration with a single 32 MHz wide IF ($\rm{432\,km\,s^{-1}}$) split into 256 channels of width $\rm{125\,kHz\approx1.7\,km\,s^{-1}}$. The sources PKS1934-638 ($\rm{0.81\,Jy}$), PKS1313-333 ($\rm{1.06\,Jy}$), and PKS1253-055 ($\rm{14.02\,Jy}$) were used as amplitude, phase and bandpass calibrators, respectively. As described before, by using the task UVLIN, a K-band continuum map from the line-free channels was produced.\\
\indent Finally, we have observed NGC\,3256 in April 2008 with the ATCA 6 km array configuration in spectral line mode at 22 GHz using the FULL\_32\_128-128 correlator configuration with one IF (Frequency 1), covering 32-MHz (432\, $\rm{km\, s^{-1}}$), all being split into 128 spectral channels of width $\rm{250\,MHz\approx3.4} \, \rm{km\, s^{-1}}$.  Simultaneously, IF2 (Frequency 2) was set to observe the continuum at 19.5 GHz with 128 MHz bandwidth. The amplitude and phase calibrators were PKS1934-638 ($\rm{0.96\, Jy}$) and PKS1004-50 ($\rm{0.80\, Jy}$), respectively. To calibrate the bandpass, we also observed PKS1921-293 ($\rm{10.45\, Jy}$). The weather was good during the entire observing session. Once again, through the MIRIAD task UVLIN, we produced a K-band continuum map from the line-free channels.

\section{Results}
Two new \hdueo \, masers were detected, in the merging system NGC\,3256 and in the spiral galaxy NGC\,3620. Line profiles and continuum maps are shown in Figs. 1-6, while line and continuum parameters are given in Tables 2 and 3.

\subsection{NGC\,3256}
\begin{figure}[h!]
\centering
\includegraphics[width= 10 cm,]{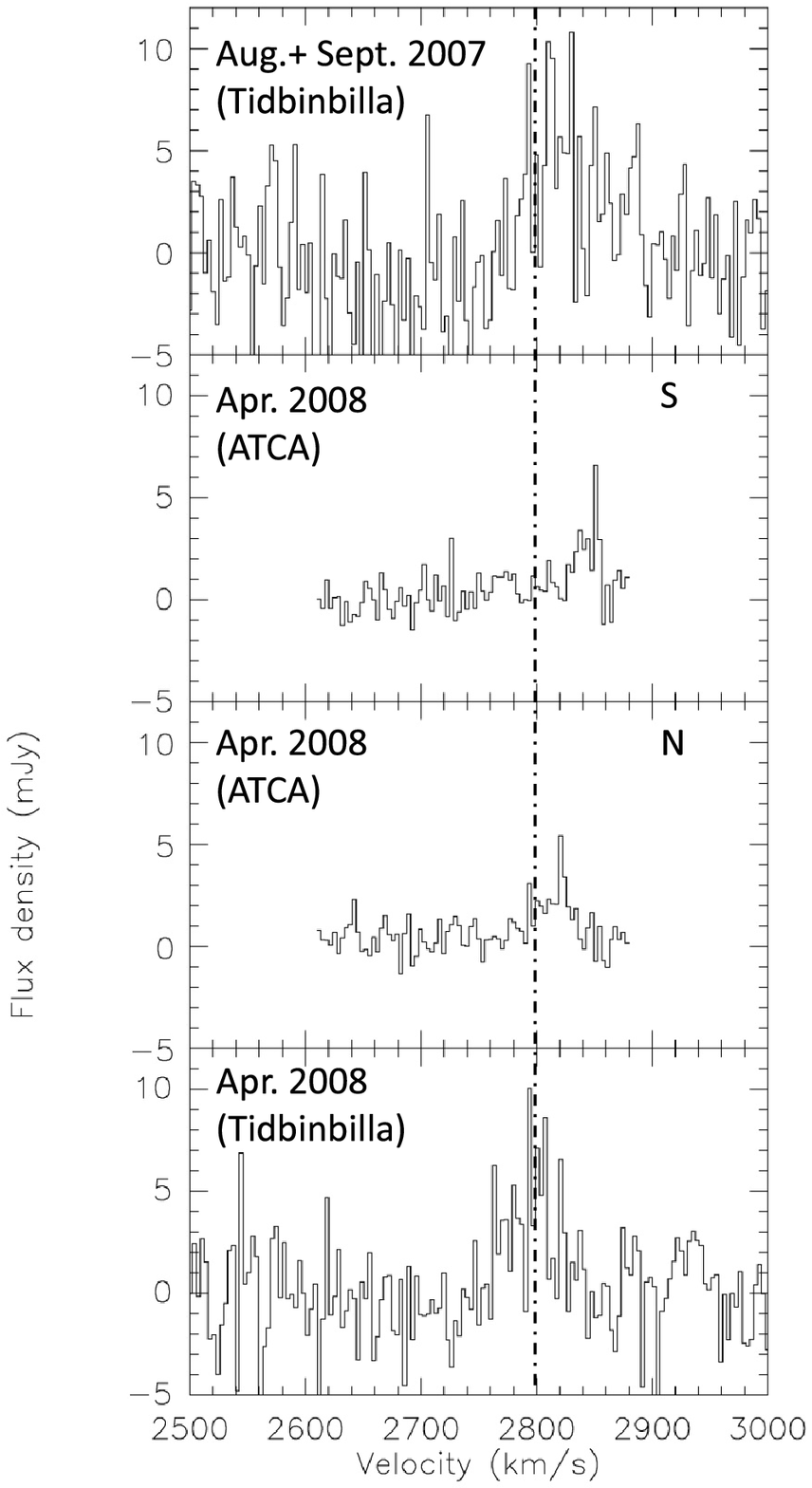}
\caption{\hdueo \, maser in NGC\,3256. The channel spacing and pointing position for the Tidbinbilla spectra are $\rm{3.4\,km\,s^{-1}}$ and $\rm{\alpha_{2000}=10^{h}27^{m}51^{s}\!.3}$, $\rm{\delta_{2000}=-43^{\circ}54'14''}$, respectively. The channel spacing for the ATCA data is $\rm{3.7\,km\,s^{-1}}$, where two emission peaks have been found at different positions (N=north, S=south). The velocity scale is with respect to the local standard of rest (LSR). The recessional LSR velocity of the galaxy (the dot-dashed line) is $\rm{2799\,km\,s^{-1}}$ according to the NASA/IPAC database (NED).}
\label{ngc3256}
\end{figure}

NGC\,3256 is a well studied merging galaxy located at a distance of $\rm{\sim37 \, Mpc}$ ($\rm{1''\sim180\,pc}$; $H_{\rm{0}}=\rm{75\, km \,s^{-1}\, Mpc^{-1}}$). With an IR luminosity of $L_{\rm{IR}}\sim6 \times 10^{11}\solum$ (Lira et al., \cite{lira08}), it is one of the most luminous galaxies in the nearby Universe. The galaxy consists of a main body, two extended tidal tails with a total extension of $\rm{\sim80\,kpc}$, and two faint external loops (L\'{\i}pari, \cite{lipari00}). The double tidal tails are characteristic of an interaction between two spiral galaxies of comparable mass (e.g., de Vaucouleurs \& de Vaucouleurs, \cite{vaun61}).\\
\indent Two distinct nuclei aligned along a north-south axis have been revealed by high spatial resolution near-IR (e.g. Kotilainen et al., \cite{Kont96}), MIR (B\"{o}ker et al., \cite{boker97}), and radio observations (e.g. Norris \& Forbes, \cite{norris95}, hereafter NF95; Neff et al., \cite{neff03}). Their projected separation is only $\rm{\sim1\,kpc}$ ($\rm{5''}$). The two nuclei have, at radio wavelengths, approximately similar sizes, brightnesses, and spectral indices. These indices are quite steep ($\rm{\alpha\sim-0.8}$), so that the radio continuum emission must be dominated by nonthermal synchrotron processes (NF95) typical of a starburst galaxy. There is abundant molecular gas ($\rm{10^{10}\,\solmass}$) in the central region surrounding the double nucleus, presumably feeding the starburst (e.g. Sargent et al., \cite{sar89}; Casoli et al., \cite{caso91}). The starburst was also detected in radio recombination lines of hydrogen and helium in the central $3''$ of the nuclear region together with prominent CO band absorption at $\rm{2.3 \mu m}$ (Doyon et al., \cite{doyon94}).\\
\indent Lira et al. (\cite{lira08}) determined a star formation rate (SFR) of $\rm{\sim15}$ and $\rm{\sim6\,\solmass yr^{-1}}$ for the northern and southern nucleus, respectively. These SFRs are in good agreement with the values obtained by NF95 using a supernova rate of $\rm{0.3\,yr^{-1}}$ per nucleus. The data presented by Lira et al. (\cite{lira08}) further suggest that a significant fraction of the mid-IR excess is due to continuum emission produced by warm dust, heated in-situ by the starburst itself.\\
\indent High-resolution spatial and spectral Chandra observations have been analysed by Lira et al. (\cite{lira02}). About $80\%$ of the X-ray emission is of diffuse origin. Each of the two nuclei is associated with one of the 14 identified discrete sources, which provide the remaining 20\% of the X-ray emission.\\
\indent Based on a comparison of radio and X-ray observations Neff et al. (\cite{neff03}) proposed the presence of a low-luminosity AGN in each nucleus, while Jenkins et al. (\cite{jen04}) found no strong evidence for an AGN in the X-ray data alone.\\
\indent We detected water maser emission in August 2007. Fig. \ref{ngc3256} shows the detected line at different epochs, while line parameters are given in Table \ref{param}. An isotropic luminosity of $\rm{\sim 10\,\solum}$ places the maser near the limit between water kilomasers and megamasers.\\
\indent In the ATCA observation, about a third of the total \hdueo \, emission is seen to originate from two separate maser spots: a northern one (labelled N) is located at $\alpha_{2000}=10^{\rm{h}}27^{\rm{m}}51^{\rm{s}}\!.87\pm0^{\rm{s}}\!.01$ and $\delta_{2000}=-43^{\circ}54'12''\!\!.4\pm0''\!\!.1$, and a southern one (labelled S) at $\alpha_{2000}=10^{\rm{h}}27^{\rm{m}}51^{\rm{s}}\!.74\pm0^{\rm{s}}\!.01$ and $\delta_{2000}=-43^{\circ}54'19''\!\!.2\pm0''\!\!.1$ (Fig. \ref{ngc3256cont}). Gaussian fits provide peak velocities of $\rm{2822\,km \,s^{-1}}$ and $\rm{2840 \,km \,s^{-1}}$, peak flux densities of $\rm{3.5\,mJy}$ and $\rm{2.1\,mJy}$, and linewidths of $\rm{8\,km\,s^{-1}}$ and $\rm{16\,km\,s^{-1}}$ for the northern and southern maser, respectively. Hence both features are redshifted with respect to the systemic velocity, by $\rm{23\,km\,s^{-1}}$ (N) and $\rm{41\,km\,s^{-1}}$ (S). The luminosities of the N and S masers are 1.0 \solum \, and 2.4 \solum, respectively.\\
\indent A few days after the ATCA observations, we observed the maser source again with Tidbinbilla. The peak velocity was $\rm{\sim2797 \,km \,s^{-1}}$, the peak flux density was $\rm{10\,mJy}$ and the linewidth was $\rm{40\,km\,s^{-1}}$, providing a single-dish luminosity of $\rm{\sim}$ 13 \solum, slightly higher than that measured in 2007.
\subsubsection{Nuclear continuum emission}
In order to associate the \hdueo \, maser with the main centers of radio emission, an interferometric radio continuum map of NGC3256 has been produced at 19.5 GHz (see Sect. 2.2). \\
\begin{figure}[h!]
\centering
\includegraphics[width = 8cm, angle=-90]{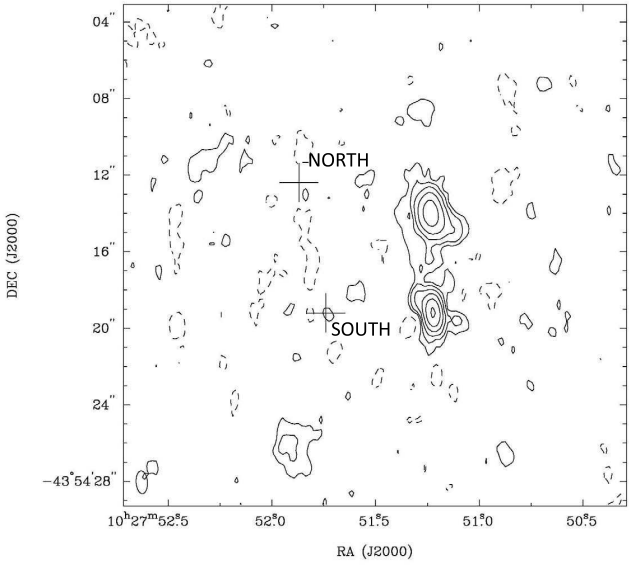}
\caption{Radio continuum map of the nuclear region of NGC\,3256, observed with ATCA at 19.5 GHz (K-band). Contour levels are $0.3 \, \rm{mJy/beam}( 3\sigma) \times -1$ (dashed contours), 1, 2, 4, 8, 16 and 32 (solid contours). For more details see Tables \ref{conttab} and \ref{mapparam}. The crosses indicate the two maser spots observed with ATCA at 22 GHz. }
\label{ngc3256cont}
\end{figure}
\indent The map (Fig. \ref{ngc3256cont}) shows the two nuclei already observed by NF95 with ATCA at 5 and 8 GHz and by Neff et al. (\cite{neff03}) with the VLA at 5, 8, and 15 GHz. In our K-band image, the projected separation between the nuclei is $5''\!\!.2\pm0''\!\!.1$, consistent with that of NF95. The emission is slightly resolved and the continuum emission of both nuclei, which is more evident in the northern one, shows an elongation toward the south-west. Weaker emission ``bridging'' the two nuclei is visible. Our map is very similar to those of NF95.\\
\indent The integrated flux density of the northern nucleus is 6.41 mJy and its position is $\alpha_{2000}=10^{\rm{h}}27^{\rm{m}}51^{\rm{s}}\!.23$, $\delta_{2000}=-43^{\circ}54'14''\!\!.0$. The southern nucleus exhibits an integrated flux of $\rm{\sim}$ 4.90 mJy at a position of $\alpha_{2000}=10^{\rm{h}}27^{\rm{m}}51^{\rm{s}}\!.22$, $\delta_{2000}=-43^{\circ}54'19''\!\!.2$ (see Table \ref{conttab}).\\
\indent NF95, based on their 5 and 8\,GHz data, reported spectral indices for the nuclei of $\rm{\alpha_{N}=-0.78}$ and $\rm{\alpha_{S}=-0.86}$. The spectral indices calculated by including our 19.5 GHz data (after a proper convolution to the 5 GHz beam) are steeper, viz. $\rm{\alpha_{N} = -1.22}$ and $\rm{\alpha_{S} = -1.05}$. They are, however, still indicative of optically thin synchrotron emission as suggested by NF95.

\subsection{NGC\,3620}
NGC\,3620 is a peculiar southern SB galaxy (Elfhag et al., \cite{elfhag96}). Located at a distance of $\rm{\sim20\,Mpc}$ ($\rm{1''\sim100\,pc}$; $H_{\rm{0}}=\rm{75\, km \,s^{-1}\, Mpc^{-1}}$), it shows a significant velocity gradient across the nucleus in optical lines ($\rm{H_{\alpha}}$, $\rm{[N{\sc II}]}$, and $\rm{[S{\sc II}]}$), which is likely caused by a rotating disk of gas (Schwartz,  \cite{schw78}). There is good agreement between the two optical velocity components, $2''\!\!.5$ east and west of the nucleus ($1649\pm9$ and $\rm{1930\pm33\,km\,s^{-1}}$, respectively; Schwartz, \cite{schw78}),
\begin{figure}[h!]
\centering
\includegraphics[width= 10 cm]{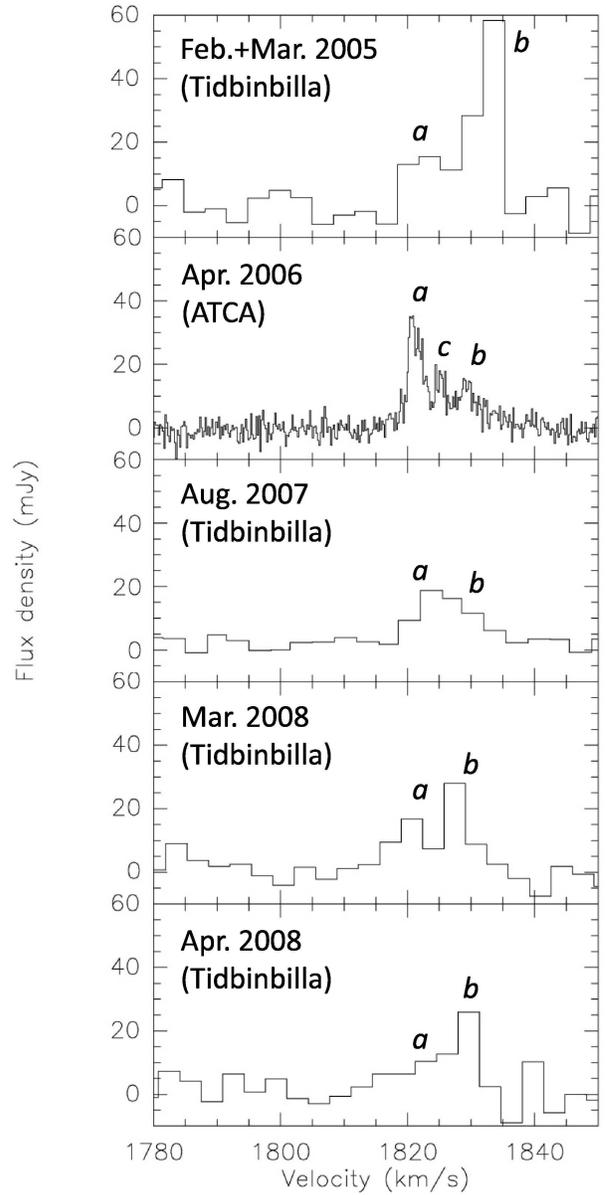}
\caption{\hdueo \, spectra observed toward NGC\,3620 between February 2005 and April 2008. Channel spacings are $\rm{3.4\,km\,s^{-1}}$ (Tidbinbilla) and $\rm{0.2 \,km\, s^{-1}}$ (ATCA). The pointing position is $\alpha_{2000}=11^{\rm{h}}16^{\rm{m}}04^{\rm{s}}\!.7$, $\delta_{2000}=-76^{\circ}12'59''$.The velocity scale is with respect to the local standard of rest (LSR). The recessional LSR velocity of the galaxy is 1691 $\rm{km\, s^{-1}}$ according to the NASA/IPAC database (NED).}
\label{ngc3620}
\end{figure}
and the values of the two major peaks in the CO profile (Elfhag et al., \cite{elfhag96}). The FIR luminosity is $\rm{\sim4\times10^{10} \, \solum}$ (Sanders et al., \cite{sand03}) and the LSR systemic velocity is $\rm{1691\,km\,s^{-1}}$. Neither OH (Staveley-Smith et al., \cite{stave92}) megamasers nor methanol masers (Phillips et al., \cite{phill98}) have been detected in NGC\,3620.

In February 2005, we detected an \hdueo \, maser in NGC\,3620 that has been monitored until April 2008. The maser emission is composed of two main spectral features (see Fig. \ref{ngc3620}), denoted ``a'' ($V_{\rm{LSR}} \rm{\sim1823 \, km \, s^{-1}}$) and ``b'' ($V_{\rm{LSR}}\rm{\sim1833 \, km \, s^{-1}}$). Both features are redshifted with respect to the systemic LSR velocity of the galaxy ($\rm{1691 \, km \, s^{-1}}$). The total isotropic maser luminosity is $\rm{\sim 4.7\, \solum}$.\\
\indent In the more sensitive ATCA observation made in April 2006, the two features, $a$ and $b$, are confirmed and a third feature (hereafter {\it c}) seems to be present between the two main ones, possibly detected because of the higher sensitivity and velocity resolution of the spectrum (see Fig. \ref{ngc3620}). The LSR peak velocity of feature $c$ is $\rm{1825 \, km \, s^{-1}}$. Details of the results obtained with these ATCA spectral observations are summarized in Table \ref{param}. All three maser components arise from a spatially unresolved spot at $\alpha_{2000}=11^{\rm{h}}16^{\rm{m}}04^{\rm{s}}\!.640\pm0^{\rm{s}}\!.003$ and $\delta_{2000}=-76^{\circ}12'58''\!\!.40\pm0''\!\!.01$. The total isotropic maser luminosity is $\rm{\sim 2.1\, \solum}$.\\
\indent While maser emission is present in all spectra, the relative intensities of the individual features vary drastically. In the 2005 spectrum feature $b$ is dominant with respect to feature $a$. This situation rapidly changes in the later spectra. In the most recent measurements the strength of feature $b$ is once again higher.\\

\subsubsection{Nuclear continuum emission}
Interferometric radio continuum maps of NGC\,3620 were, so far, not yet reported. Hence, in order to 
\begin{figure}[h!]
\centering
\includegraphics[width = 9cm]{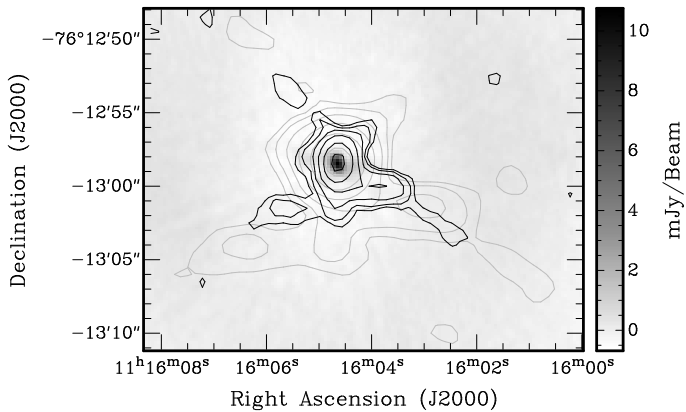}
\caption{Radio continuum maps of the nuclear region of NGC\,3620, observed with ATCA. C-band (4.8 GHz) emission is shown in grey contours, X-band (8.6 GHz) emission  in black contours, and K-band (19.5 GHz) emission, confined toward the nucleus, in grey scale. C- and X-band contour levels are $3\sigma \times -1, 1, 2, 4, 8, 16, 32, 64$ with $\sigma$ = 0.4 and 0.2 mJy/beam, respectively. For more details see Tables \ref{conttab} and \ref{mapparam}. }
\label{ngc3620cont}
\end{figure}
improve our understanding of the central region of the galaxy and, possibly, to associate the newly detected \hdueo \, maser with a putative nucleus, we obtained radio continuum maps at three frequencies. Figure \ref{ngc3620cont} shows the ATCA images of the central region taken at 4.8, 8.6 and 19.5 GHz (details of these maps are given in Tables \ref{conttab} and \ref{mapparam}).\\
\indent Peak intensity positions at the different wavelengths all coincide at $\alpha_{2000}=11^{\rm{h}}16^{\rm{m}}04^{\rm{s}}\!.7$ and $\delta_{2000}=-76^{\circ}12'58''\!\!.33$. The C-band emission (Fig. \ref{ngc3620cont}, in grey contours) is resolved into a compact nuclear part and some diffuse emission stretching toward southwestern and southeastern directions. A similar morphology is also observed at 8.6\,GHz. K-band emission (19.5 GHz, in grey scale), in contrast, is more compact.\\
\indent A K-band (22 GHz) continuum map from the ATCA data cube has been produced using only the line-free channels. The position of the peak emission ($\rm{\sim7 \,mJy \,beam^{-1}}$) is coincident with the peak of the broad-band ATCA continuum map at 19.5 GHz (Fig.\ref{ngc3620cont}) and, more important, coincident with that of the maser spot within an overall positional accuracy of $\rm{\sim 1''\!\!}$. This was obtained from the quadratic sum of the uncertainties in our line and continuum positions.
\begin {table}[h]
\caption []{Gaussian fit parameters of the continuum maps.} 
\begin{center}
\scriptsize
\begin{tabular}{l l c c c c }
\hline
\hline
\\
Galaxy & Band   & peak flux      & integrated  & rms       \\
       &        & density         & flux density  &           \\
       &        & (mJy/beam)& (mJy)           & (mJy)\\
\\
\hline
\\
NGC3256 (N)& K   & $2.36\pm0.12$     & 6.41            &  0.1\\
\; \qquad \qquad (S)& & $3.37\pm0.18$& 4.90            &  \\
           &     &                 &                 &\\
NGC3620& C       & $45.96\pm0.60$    & 73.94$^{a}$     &  0.4 \\
       & X       & $21.48\pm0.29$    & 49.02$^{a}$     &  0.2 \\
       & K       & $10.05\pm0.30$    & 29.34$^{a}$     &  0.2\\
       & K$^{b}$ & $7.12\pm0.27$    & 16.81$^{a}$     &  0.2\\
\\
\hline
\end{tabular}
\end{center}
\scriptsize{$^{a}$ Obtained by Gaussian fits of maps restored using the beam at 4.8 GHz\\
$^b$ Obtained by subtracting line channels from spectral data at 22 GHz.}
\label{mapparam}
\end{table}
\section{Discussion}
\subsection{The nature of the masers in NGC\,3256 and NGC\,3620}
The southern sample ($\rm{Dec<-30^{\circ}}$) targets with FIR $\rm{100\,\mu m}$ flux density $\rm{S_{100 \, \mu m}>50 \, Jy}$ is composed of 20 galaxies. Water maser emission has been detected in three galaxies: in NGC\,4945 (Dos Santos \& L\`{e}pine, \cite{santos79}) and in NGC\,3256 and NGC\,3620 (Sect. 3).\\

\begin{figure}[h!]
\centering
\includegraphics[width= 8 cm, angle=-90]{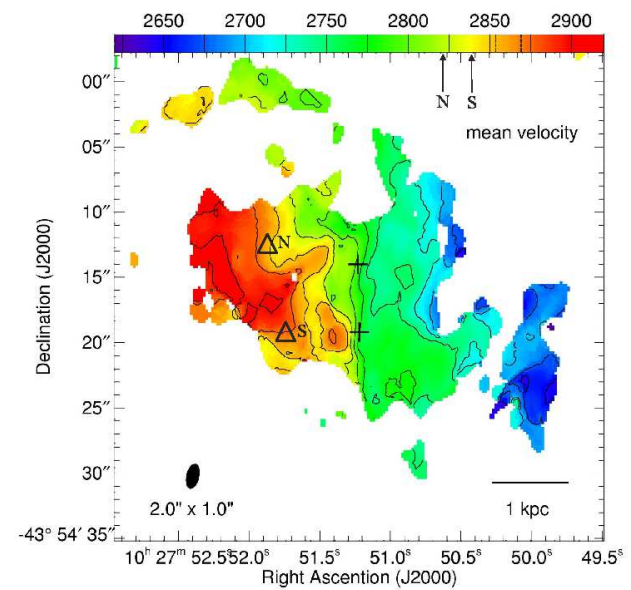}
\caption{High-resolution $^{12}$CO\,(2-1) map of the mean velocity in units of km s$^{-1}$ toward the nuclear region of NGC\,3256 (Sakamoto et al. \cite{saka06}, their Fig. 4b). Triangles indicate the positions of the water maser spots found by ATCA (Fig. \ref{ngc3256cont}) and the crosses indicate the positions of the two nuclei.}
\label{saka4b}
\end{figure}

\begin{figure}[h!]
\centering
\includegraphics[width=8 cm]{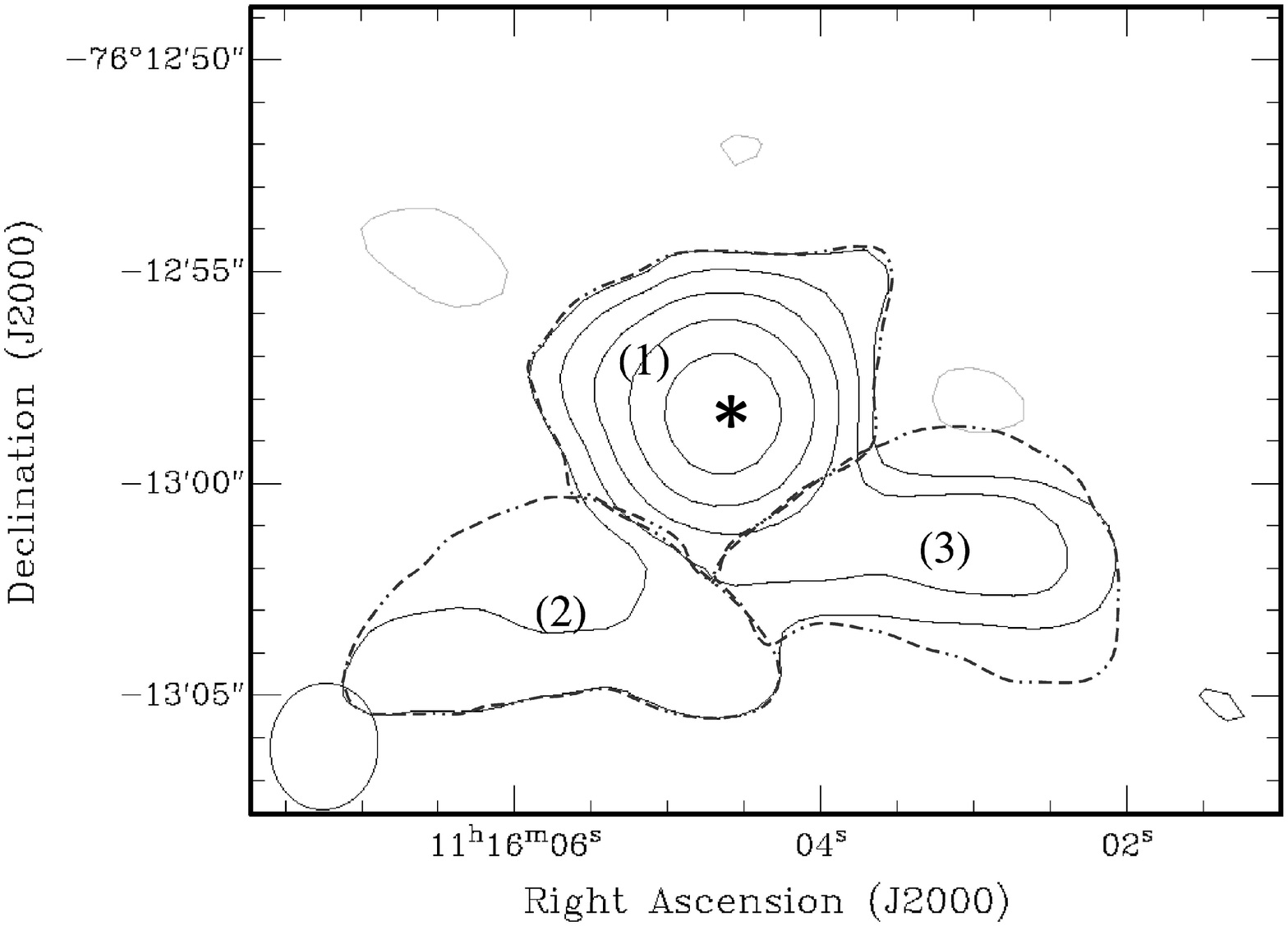}
\caption{Radio continuum map of the nuclear region of NGC\,3620 at 4.8 GHz. Contour levels are $1.2 \,\rm{mJy/beam} \times -1$ (grey contours) , 1, 2, 4, 8, 16, 32 (black contours). The asterisk (in the center of the map) marks the ATCA position of the \hdueo \, kilomaser. Three regions are indicated by dash-dash-dot-dotted lines: the central emission (1), the southeastern emission (2) and the southwestern emission (3). The beam is shown at the lower left corner.}
\label{maspos}
\end{figure}

\paragraph{NGC\,3256:} The $L_{\rm{H_{2}O}}\sim11\, \solum$ maser emission detected at Tidbinbilla is resolved by ATCA into two spots of luminosity 1.0 \solum \, (N) and 2.4 \solum (S) (Figs. \ref{ngc3256} and \ref{ngc3256cont}). Since the sum of the luminosities of the two maser hotspots is not sufficient to account for the single dish flux, the single-dish emission must be the result of a superposition of several undetected maser hotspots of which only the two strongest have been detected at high resolution. This is a situation similar to that encounterd in NGC\,2146 by Tarchi et al. (\cite{tarchi02}).\\
\indent Sakamoto et al. (\cite{saka06}) studied the gas kinematics around each nucleus at high-resolution. Comparing our maser velocities with those of their $^{12}$CO\,(2-1) map, we find agreement with the galaxy kinematics (Fig. \ref{saka4b}). Both masers are located eastward of the double nucleus and are red-shifted with respect to the systemic velocity (e.g., English et al., \cite{eng03}). When the velocities of the two maser lines are compared with those of the neighbouring CO gas, the discrepancies are of order 25 km s$^{-1}$,  which can be justified in terms of local motions of the masing gas. Hence, the \hdueo \, kilomasers in NGC\,3256 are not associated with either one of the two nuclei but with the main gas disk hosting an ongoing starburst (Sakamoto et al., \cite{saka06}).
\paragraph{NGC\,3620:} The kilomaser emission in NGC\,3620 is close to the nuclear region of the galaxy (indicated by an asterisk in Fig. \ref{maspos}). The swap of the two main line features between 2005 and 2006 and the subsequent swap-back in the following epochs let us speculatively think about an anticorrelation between the maser features  associated with an accretion disk like the one in the galactic star forming region S255 (Cesaroni, \cite{cesa90}). The association of the kilomaser with either a star forming region or the nucleus of the galaxy requires a search for a so far undetected Low Luminosity AGN (see Sect. 4.1.1).
\subsubsection{What is the nuclear source in NGC\,3620?}
\indent The spectral index was computed from the three ATCA maps produced at 4.8, 8.6 and 19.5 GHz by fitting the three points in the diagram ($\rm{\nu,\,S_{\nu}}$), using the convention $\rm{S \propto \nu^{\alpha}}$.
The resolution of the maps is different and ranges from $\sim2$ arcsec to $\sim0.5$ arcsec. Therefore, a convolution to the same 2 arcsec beam (that at 4.8 GHz) was necessary. The computed spectral indices derived from peak and integrated flux densities (see Table \ref{mapparam}) are $-1.07$ and $-0.65$, respectively. These values indicate that the radio emission is most likely dominated by optically thin synchrotron emission typical for a collection of sources related to star formation activity like radio supernovae (RSNs) and supernova remnants (SNRs). Hence, the nuclear region of NGC\,3620 is linked to intense star formation, and, apparently, there is no evidence for an AGN (whose spectral index may be flatter) in the center of the galaxy. However, the spectrum might be artificially steepened by missing flux at the higher frequencies. Furthermore, when the nuclear radio continuum emission is divided into three sub-regions, (1) the central emission, (2) the southeastern emission, and (3) the southwestern emission, the spectral indices, computed by using the integrated flux densities, of the three regions differ slightly. In particular, region (1) has a flatter spectral index ($\alpha_{1}$ = --0.64) than those of regions (2) and (3) ($\alpha_{2}$ = --2.22 and $\alpha_{3}$ = --1.54, respectively). While this is not unusual and can be justified in several other ways, the possibility exists that this is caused by a nuclear component 
with flatter or inverted spectral index. Moreover, since the angular resolution of our observation is low ($2''$ correspond to 200 pc), the lower limit to brightness temperature in region (1) at X-band ($T_{\rm{B}}\geq\rm{10^{2} \, K}$) does not allow us to distinguish between a thermal or non-thermal nature of the central region. Continuum Long Baseline Array (LBA) observations of NGC\,3620 at 1.6 GHz have been accepted and await scheduling. This aims at (1) identifying nuclear radio continuum sources with particularly high brightness temperature, (2) locating with higher accuracy the radio continuum emission with respect to the maser spot and hence, (3) establishing if the maser is associated with star formation like the majority of kilomasers or, if present, with AGN activity.

\subsection{Statistical considerations}
\indent The detection rate in the southern sample is 3/20 or, applying the Bernoulli theorem, ($15\pm$8)\%. As the rms in the northern and southern samples is comparable (see Table \ref{maser}) because of different integration times, this detection rate is compatible with that obtained by HPT, i.e ($22\pm$6)\%. Very recently, Braatz \& Gugliucci (\cite{braatz08}) have found a water kilomaser in NGC\,4527, which remained undetected by HPT. Darling et al. (\cite{darl08}) have also reported the detection of a new water kilomaser in the Antennae system that is composed of two galaxies, NGC\,4038 and NGC\,4039. This demonstrates that the high sensitivity of the Green Bank Telescope (GBT) can further increase the number of detected kilomasers in our sample (see below). With these new detections the northern detection rate becomes ($27\pm$7)\% (12/45).\\
\indent Adding the southern sources, the complete all-sky sample with  $\rm{S_{100 \, \mu m}>50 \, Jy}$ is finally composed of 65 galaxies (Table \ref{maser}). Among them, water maser emission has been detected in 15 galaxies. There are 10 kilomasers (IC\,10, NGC\,253, IC\,342, NGC\,2146, M\,82, NGC\,3556, NGC\,3620, NGC\,4038/9, NGC\,5194, and NGC\,4527), 4 megamasers (NGC\,1068, Arp\,299, NGC\,3079, and NGC\,4945) and one maser (NGC\,3256) with an isotropic luminosity just above the 10 \solum \,threshold conventionally used to discriminate between megamasers and kilomasers. Since the maser emission is associated with sites of massive star formation and since individual ATCA maser spots show luminosities well below the 10 \solum \, threshold, NGC\,3256 is also a kilomaser source. As already mentioned in Sect.\,1, the majority of the kilomasers are associated with star formation. Most of them show a collection of Galactic-type masers related to (pre-) stellar activity. Therefore, a relation between kilomasers and FIR emission is readily explained, since the latter commonly arises from warm dust grains heated by young massive stars. The bulk of the 22 GHz \hdueo \, emission from the four megamasers in our sample (NGC\,1068, NGC\,3079 and NGC\,4945 but presumably not Arp\,299) is associated with nuclear accretion disks surrounding an AGN (Gallimore et al., \cite{galli97}; Greenhill, Moran \& Herrnstein, \cite{green97}; Trotter et al., \cite{tro98}; Tarchi et al., \cite{tarchi07}).\\
\indent In our complete sample, 40\% (6/15) of the water masers have been detected in galaxies classified as Seyfert, all of type 2 (see Table \ref{maser}). The other 9 masers are detected in spiral galaxies (27\%, 4/15), merger systems (20\%, 3/15), LINERs (7\%, 1/15), and the last one in a starburst galaxy (7\%, 1/15).\\
\indent The high rate of maser detections in the complete all-sky sample (23\%, 15/65) confirms the link between FIR flux density and maser phenomena as previously suggested by HPT and CTH. In particular, the detection rate  strongly declines with decreasing FIR flux. Adding the masers detected in the southern sample to those of HPT, for fluxes of  $\rm{300-1000\,Jy}$, $\rm{100-300\, Jy}$, and $\rm{50-100\,Jy}$, we find detection rates of 3/4 or 75\%, 6/21 or 29\%, and 6/40 or 15\%.\\
\begin{figure}[h!]
\centering
\includegraphics[width=9 cm]{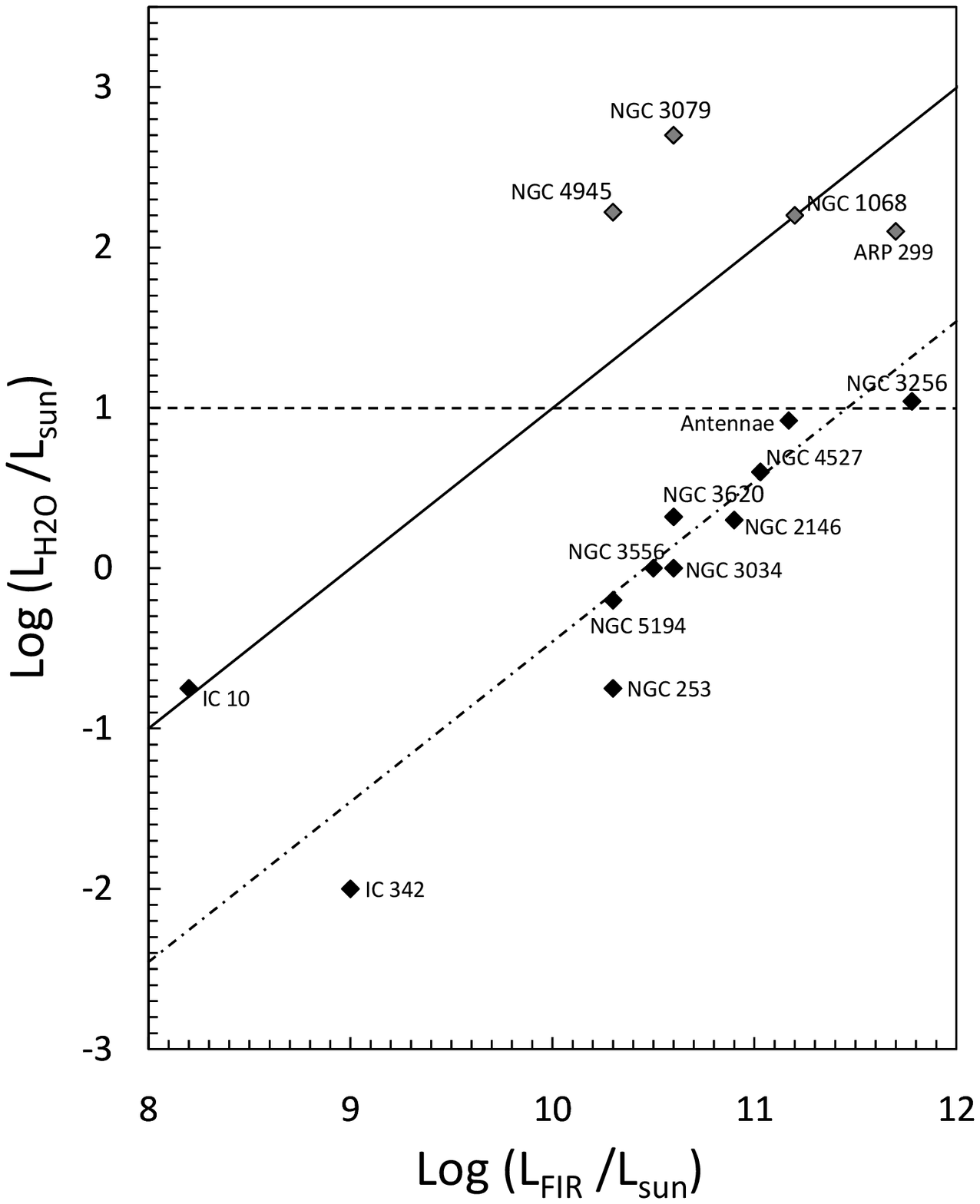}
\caption{\hdueo \, vs. FIR luminosity including the 15 masers of the complete all-sky FIR-sample, and the line separating between the kilomaser and megamaser regimes for \hdueo \, masers (dashed horizontal line). The solid line shows the correlation found by Jaffe et al. (\cite{jaf81}) for Galactic sources (log$(L_{\rm{H_{2}O}})=$log$(L_{\rm{FIR}})-9$). The dot-dashed diagonal line shows the FIR/\hdueo \, luminosity relation for the kilomaser sources (log$(L_{\rm{H_{2}O}})=$log$(L_{\rm{FIR}})-10.4$), with the slope being fixed to that of the Jaffe et al. (\cite{jaf81}) correlation. Black diamonds denote kilomasers, grey diamonds mark megamasers. The figure is a modified version of Fig.6 of CTH.}
\label{FIR_maser}
\end{figure}
\indent HPT found a proportionality between $L_{\rm{FIR}}$ and $L_{\rm{H_{2}O}}$ for extragalactic water masers, which is surprisingly consistent with that obtained for the orders of magnitude less luminous Galactic sources of Jaffe et al. (\cite{jaf81}). Furthermore, CTH observed that megamaser and kilomaser sources seem to follow independent correlations (the two samples diverge by $\rm{\sim 1.5}$ orders of magnitude in maser luminosity for a given FIR luminosity), with the tighter one being that for kilomasers. In Fig. \ref{FIR_maser}, we plot the values for the masers of the complete all-sky FIR-sample including the known maser in NGC\,4945 and the most recent detections in NGC\,3256, NGC\,3620, NGC\,4527 ($L_{\rm{FIR}}\sim1 \times 10^{11} \, \solum$) and the Antennae ($L_{\rm{FIR}}\sim 1.5 \times 10^{11} \, \solum$). The values for the luminosities of the kilomasers in our all-sky sample follow the trend given by log$(L_{\rm{H_{2}O}})=$log$(L_{\rm{FIR}})-10.4$. The most discrepant point is that representing IC\,10, which is also peculiar in the number of water masers as recently pointed out by Brunthaler et al. (\cite{bru06}). Overall, the water kilomasers in our all-sky FIR-sample reinforce the $L_{\rm{FIR}}/L_{\rm{H_{2}O}}$ correlation for kilomasers found by CTH. 

\subsubsection{GBT and EVLA: how many more masers could be detected in FIR-sample?}
The entire sample of FIR bright ($\rm{S_{100 \, \mu m}>50 \, Jy}$) galaxies was observed with three different telescopes, the 100-m Effelsberg, the 70-m Tidbinbilla, and the 100-m Green Bank telescope (GBT). Each telescope has a different sensitivity at 22 GHz, i.e. 0.83 (www.mpifr-bonn.mpg.de/div/effelsberg/calibration), 0.67 (www.atnf.csiro.au/observers/tidbinbilla) and 1.50 K/Jy (Minter \cite{min09}), using an antenna temperature scale ($\rm{T_{A}^{*}}$), with respect to a point source. For this study, the smaller size of the Tidbinbilla telescope was compensated by longer integration times with respect to the Effelsberg telescope (see Table \ref{maser}).\\
\indent Future observations with higher sensitivity with the GBT and EVLA will likely lead to new detections. It is therefore of interest to evaluate this increase in the number of water masers. The evaluation can be made considering the standard equation for the water maser density as reported by HPT for a specific slope of the luminosity function ($\gamma_{SLF}$, about --1.5, see below), that is
\begin{equation}
N_{\rm{H_{2}O}}=C_{1}L^{\rm{\gamma_{SLF}}}_{\rm{H_{2}O}}.
\end{equation}
\noindent
The integration of eq.(2) from a minimum luminosity, which is due to the sensitivity of the radiotelescope, to infinity gives
\begin{equation}
N_{\rm{tot}}=-\frac{C_{1}}{\rm{\gamma_{SLF}+1}} L^{\rm{\gamma_{SLF}+1}}_{\rm{min}}=C_{2} \, L^{\rm{\gamma_{SLF}+1}}_{\rm{min}}.
\end{equation}
\noindent Rewriting eq.(3) for the GBT and the Effelsberg/Tidbinbilla (ET) radiotelescopes, we obtain
\begin{equation}
N_{\rm{ET}}=C_{2} \, L^{\rm{\gamma_{SLF}+1}}_{\rm{min,ET}}, \qquad N_{\rm{GBT}}=C_{2} \, L^{\rm{\gamma_{SLF}+1}}_{\rm{min,GBT}}.
\end{equation}
\noindent With $\tau$ being the ratio between the ET and the GBT sensitivity, we then obtain from eqs. (4)
\begin{equation}
N_{\rm{GBT}}=C_{2} \, (\tau)^{\rm{\gamma_{SLF}+1}} \, \,  L^{\rm{\gamma_{SLF}+1}}_{\rm{min,ET}}.
\end{equation}
\noindent The ratio of the GBT and ET total number of detectable masers is
\begin{equation}
\frac{N_{\rm{GBT}}}{N_{\rm{ET}}}=(\tau)^{\rm{\gamma_{SLF}+1}}\simeq1.4,
\end{equation}
where we have assumed $\tau = 0.55$ and $\rm{\gamma_{SLF}}$ = --1.5 (HPT, Bennert et al., \cite{ben09}). Assuming that the EVLA sensitivity will be 2.44 K/Jy (Perley et al., \cite{per06}; www.gb.nrao.edu/gbt),
\begin{equation}
\frac{N_{\rm{EVLA}}}{N_{\rm{ET}}}\simeq1.7.
\end{equation}
\indent As the number of masers in the FIR-sample is 13 (considering only those detected using the ET radiotelescopes), performing a survey on the same sample with the GBT and the EVLA, with an observing time comparable with that used in our survey with Effelsberg, we can expect to obtain 5 and 9 new maser detections, respectively. Two new detections in sources belonging also to our sample have indeed recently been reported by Braatz \& Gugliucci (\cite{braatz008}) and Darling et al. (\cite{darl08}). Although the integration times in these surveys with the GBT were shorter than in our Effelsberg ones, their GBT detection thresholds were better confirming the usefulness of deeper surveys. Hence, in that case the detection rate for the FIR-sample is expected to rise to 28\% and 34\%, respectively.

\subsection{Is there a relation between OH and \hdueo \, maser emission?}
\indent Out of the 65 sources of our all-sky FIR sample, for 45 galaxies (i.e. $\sim$70\%) searches for $\lambda = 18 \, \rm{cm}$ hydroxyl (OH) maser emission have been reported in literature. In 6 galaxies OH maser emission was detected (i.e. $\sim$13\%). The percentage of galaxies in our sample that have not been searched for OH masers is higher at lower FIR fluxes (about 30\% and 40\% for $\rm{S_{100 \, \mu m}}$ below 200 and 100 Jy, respectively).\\
\begin{figure}[h!]
\centering
\includegraphics[width=9 cm]{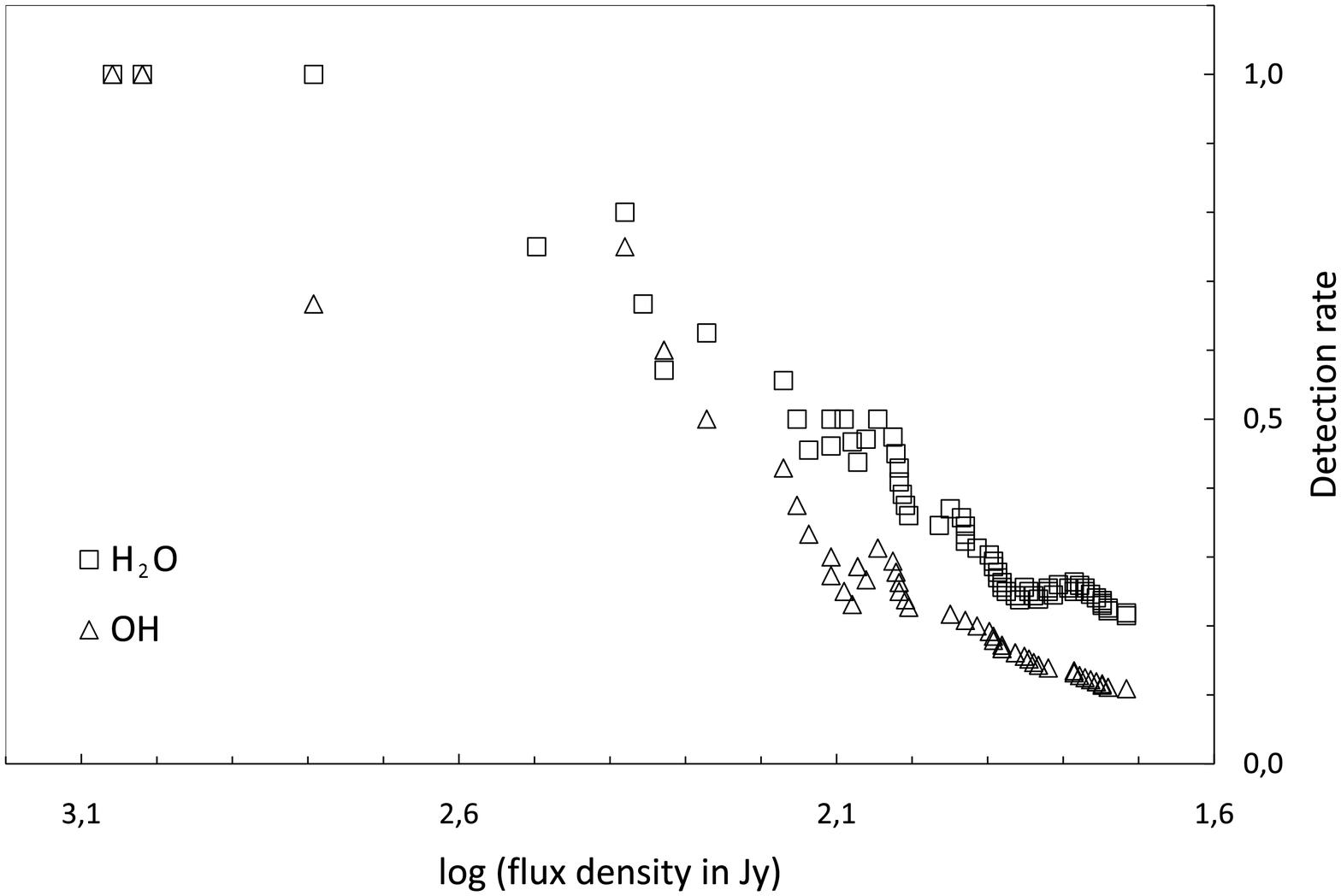}
\caption{Total \hdueo \, and OH maser cumulative detection rates above a given IRAS Point Source Catalog $\rm{S_{100 \, \mu m}}$ flux density. The figure is an expanded version of Fig.13 of HPT, with the data from OH and our sample being included.}
\label{cumdet}
\end{figure}
\indent In Fig. \ref{cumdet}, both the \hdueo \, (squares) and OH (triangles) total cumulative detection rates above a given $\rm{100\mu m}$ IRAS Point Source Catalog flux are plotted. {The trends are similar. Naturally, these trends also reflect the fact that the surveys involved are sensitivity-limited and have, for the two maser species, different detection thresholds. For \hdueo \, masers, the influence produced by a possible distance bias has been extensively discussed in HPT (their Sect. 4.3). \\
\begin{figure}[h!]
\centering
\includegraphics[width=8 cm]{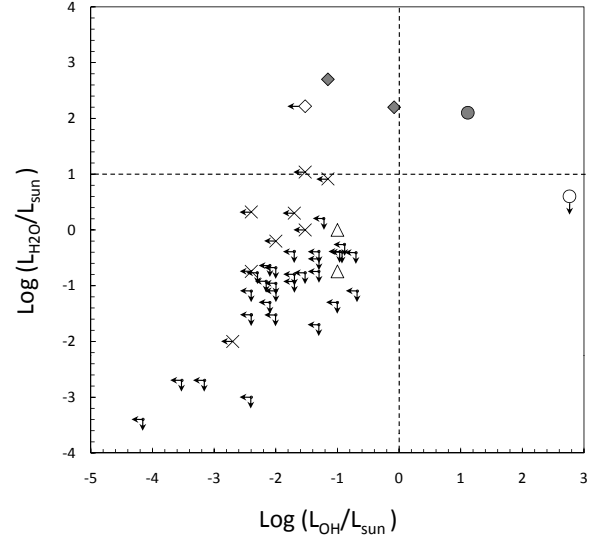}
\caption{OH vs. \hdueo \, maser luminosity plot including all sources belonging to the complete all-sky FIR-sample where masers from both species have been searched for. Crosses represent the galaxies hosting only \hdueo \, kilomasers, triangles both \hdueo \, kilomasers and weaker OH masers, diamonds \hdueo \, megamasers, filled diamond both \hdueo \, megamaser and weaker OH maser,  circles OH megamasers and the filled circle both \hdueo \, and OH megamasers. For non detections, the arrow marks the detection luminosity threshold ($3\sigma$) of the survey in that maser transition. The dashed lines indicate the separation between kilomaser and megamaser regimes as found in literature (for \hdueo \,masers, see Sect.\ 1; for OH masers, see Henkel \& Wilson \cite{hen90}).}
\label{H2O}
\end{figure}
\indent In Fig. 9, we plot maser isotropic luminosities for all sources (45 out of 65) of the all-sky FIR-sample where \hdueo \, and OH masers have been searched for (see also Table \ref{maser}). Arrows indicate upper limits for \hdueo \, and/or OH maser luminosity in case of non detections (3$\sigma$ noise level). For example, water maser emission was detected in NGC\,4945 with 165 \solum \, (Dos Santos \& L\`{e}pine, \cite{santos79}), while it remains undetected, so far, in $\lambda = 18 \, \rm{cm}$ OH down to a detection threshold of 0.46 \solum (Staveley-Smith et al., \cite{stave92}). Among the 45 sources, no detection in both maser species has been reported for 31 sources, 14 are \hdueo \, detections, and 6 host OH maser emission. Both maser species have been detected in five galaxies: three water megamaser sources, NGC\,1068, NGC\,3079, and Arp\,299, and two water kilomaser sources, the well-known starburst galaxies NGC\,253 and NGC\,3034 (M82). \\
\indent In Galactic star forming regions, the two maser species are very often associated suggesting that the OH and \hdueo \, masers occur near a common energy source but in physically distinct zones (e.g. Forster \& Caswell, \cite{for89}). This latter fact is likely a consequence of the diverse excitation and pumping conditions of the respective lines. OH maser emission, excited mostly radiatively and requiring comparatively low density gas, is more extended and farther off the excitation source than the water masers that are likely collisionally excited and require higher temperature and density (e.g. Lo \cite{lo05}; Tarchi et al., \cite{tarchi07}). Typical angular separations between the two maser group associations in Galactic star-forming regions are a few arcsecconds corresponding to linear separations of $\lesssim$\,1 parsec at a distance of $\lesssim$\,10 kpc. Hence, with the relatively coarse resolution of our single-dish surveys, masers of the two species associated with star formation would be seen, if present, in the same beam. This seems indeed to be the case for our sample.\\
\indent If we consider that six galaxies are detected in OH and assuming that \hdueo \, and OH masers are unrelated phenomena within the sample being considered, we can compute the expected number of those galaxies to be detected in \hdueo \,. This number is $\sim$ 1, while, instead, the number in our sample of sources showing both masers is 5. Furthermore, for 31 galaxies, where no OH maser has been detected, also \hdueo \, maser emission is not seen. Therefore, we are inclined to conclude that, similar to Galactic sources, an association between \hdueo \, and OH emission does exist also for extragalactic targets. Noticeably, with the only exception of Arp\,299 (Tarchi et al., \cite{tarchi07}), the contemporary presence of megamaser emission\footnote{The isotropic luminosity typically adopted to discriminate between OH mega and kilomaser sources is 1 \solum \,(Henkel \& Wilson \cite{hen90}) while it is 10 \solum \,for \hdueo \, masers (see Sect. \ref{intro}).} from both species has never been reported. This may indicate that while weaker, Galactic-type, \hdueo \, and OH masers are often associated (if not correlated) like in our Galaxy, the most luminous masers of the two species seem to be mutually exclusive.\\
\indent Given the relatively low number of sources involved, we are aware of the limited statistical relevance of this result. Furthermore, statistics may be affected by the different and/or insufficient sensitivities of maser searches and by possible variability in the maser features. In order to provide more statistically-relevant support to our arguments, a more extensive study of an entire sample comprised of all the detected extragalactic water ($\sim$100; Braatz et al., \cite{braatz07}) and/or hydroxyl maser sources ($\sim$100; Darling, \cite{darl07}) is recommendable. This is, however, outside the scope of the present work and will be the subject of another publication.

\begin {table*}[t]
\caption []{The complete all-sky FIR sample.}
\begin{center}
\scriptsize
\begin{tabular}{l l c c c c c c c c}
\hline
\hline
\\
Source$^{a}$&Classification$^{b}$& R.A.$^{b}$ & Dec.$^{b}$ &$V_{sys}^{b}$&$S_{100\mu m}^{c}$ &rms&$L_{\rm{H_{2}O}}^{d}$ &$L_{\rm{OH}}^{d}$&Ref.$^{e}$\\ 
         &                    &            &             &              &           &         &           &   &\\
         &                    & (J2000)    & (J2000)     &   (km/s)     &  (Jy)     &  (mJy)  &   (\solum)& (\solum)  &\\
\\
\hline
\\
NGC\,0055 & SB(s)m:sp         & 00h14m53.6s & -39d11m48s &  +149        & 78        &  10   & $<$0.001& $<$0.004 & w1, h1\\
\bf{IC\,10} & dIrr IV/BCD       & 00h20m17.3s & +59d18m14s &  -358        & 71      &  16   & 0.18    & $<$0.004 & w2, h2\\
NGC\,0134 & SAB(s)bc HII      & 00h30m22.0s & -33d14m39s &  +1598       & 57        &  5    & $<$0.08 & $<$0.21 & w2, h1\\
\bf{NGC\,0253}& SAB(s)c;Sbrst Sy2 HII & 00h47m33.1s& -25d17m18s& +251       & 1045  &  10   & 0.18    & 0.1 & w2, h3, h6\\
NGC\,0300 & SA(s)d            & 00h54m53.5s & -37d41m04s & +152        & 75         &  9    & $<$0.001&   -     & w1\\
NGC\,0660 & SB(s)a pec; HII LINER & 01h43m02.4s & +13d38m42s & +838     & 104       &  10   & $<$0.07 &   -     & w2 \\
NGC\,0891 & SA(s)b? sp HII    & 02h22m33.4s & +42d20m57s & +548         & 148       &  20   & $<$0.08 & $<$0.004 & w2, h2\\
NGC\,0972 & Sab HII           & 02h34m13.4s & +29d18m41s & +1555        & 65        &  10   & $<$0.65 &   -     & w2\\
NGC\,1055 & SBb; sp; Sy2  LINER & 02h41m45.2s& +00d26m35s& +989        & 60         &  17   & $<$0.23 & $<$0.008 & w2, h4\\
Maffei 2  & SAB(rs)bc:        & 02h41m55.1s & +59d36m15s & -23          & 227       &  17   & $<$0.0001&   -    & w2\\
\bf{NGC\,1068}& SA(rs)b;Sy1 Sy2& 02h42m40.7s & -00d00m48s & +1131        & 240      &  40   &  158    &  0.83 & w2, h5\\
NGC\,1084 & SA(s)c HII        & 02h45m59.9s & -07d34m43s & +1427        & 55        &  20   & $<$0.55 & $<$0.13 & w2, h2\\
NGC\,1097 & SB(r'1)b Sy1& 02h46m19.0s& -30d16m30s & +1253       & 85                &  15   & $<$0.30 & $<$0.05 & w1, h6\\
NGC\,1313 & SB(s)d HII        & 03h18m16.0s & -66d29m54s & +456         & 92        &  4    & $<$0.005&     -   & w1\\
NGC\,1365 & SBb(s)b;HII Sy1.8 & 03h33m36.4s &-38d08m25s & +1616     & 142           &  12   & $<$0.39 & $<$0.20& w1, h1\\
\bf{IC\,342}& SAB(rs)cd; Sy2 HII& 03h46m48.5s & +68d05m46s & +49          & 128     &  12   &  0.01   & $<$0.002 & w2, h3\\
UGC\,02855& SABc              & 03h48m20.7s & +70d07m58s & +1186        & 79        &  9    & $<$0.17 & $<$0.03 & w2, h2\\
NGC\,1559 & SB(s)cd           & 04h17m35.8s & -62d47m01s & +1307        & 56        &  5    & $<$0.05 & $<$0.10 & w1, h1\\
NGC\,1569 & IBm;Sbrst Sy1     & 04h30m49.0s & +64d50m53s & -122         & 52        &  12   & $<$0.002& $<$0.0007 & w2, h2\\
NGC\,1672 & SB(r)bc Sy2& 04h45m42.5s & -59d14m50s & +1339       & 70                &  19   & $<$0.41 & $<$0.11 & w1, h1\\
NGC\,1792 & SA(rs)bc          & 05h05m14.4s & -37d58m51s & +1221        & 77        &  5    & $<$0.05 &    -    & w1\\
NGC\,1808 & SAB(s:)b Sy2& 05h07m42.3s & -37d30m47s & +995        & 137              &  15   & $<$0.18 & $<$0.05 & w1, h7\\
\bf{NGC\,2146}& SB(s)ab pec HII   & 06h18m37.7s & +78d21m25s & +891     & 187       &  7    &  2      & $<$0.02 & w2, h2, h6\\ 
NGC\,2403 & SAB(s)cd HII      & 07h36m51.4s & +65d36m09s & +123         & 56        &  9    &$<$0.002 & $<$0.0003& w2, h2\\
NGC\,2559 & SB(s)bc pec:      & 08h17m06.1s & -27d27m21s & +1576        & 66        &  20   & $<$0.67 &    -    & w2\\
NGC\,2903 & SB(s)d HII        & 09h32m10.1s & +21d30m03s & +553         & 104       &  7    &  $<$0.03& $<$0.01 & w2, h2, h6\\
NGC\,2997 & SA(s)c            & 09h45m38.8s & -31d11m28s & +1081        & 85        &  7    &  $<$0.05&    -    & w1\\
\bf{NGC\,3034} (M82) & I0;Sbrst HII& 09h55m52.7s & +69d40m46s & +287    & 1145      &  20   &  1      & 0.1 & w2, h3, h6\\
\bf{NGC\,3079} & SB(s)c;LINER Sy2  & 10h01m57.8s & +55d40m47s & +1128   & 89        &  40   &  501    & 0.07     & w2, h8\\
\bf{NGC\,3256}& Pec;merger;HII Sbrst & 10h27m51.3s & -43d54m14s & +2799 & 115       &  7    &  10.9   & $<$0.02 & w1, h1\\
NGC\,3521 & SAB(rs)bc;HII LINER &11h05m48.6s & -00d02m09s & +795        & 85        &  29   &  $<$0.25&    -    & w2\\
\bf{NGC\,3556} & SB(s)cd HII       & 11h11m31.0s & +55d40m27s & +708    & 61        &  10   &  1      & $<$0.03 & w2, h6\\
\bf{NGC\,3620} & SB(s)ab           & 11h16m04.7s & -76d12m59s & +1691   & 66        &  7    &  2.1    & $<$0.004 & w1, h7\\
NGC\,3621 & SA(s)d HII        & 11h18m16.5s & -32d48m51s & +749         & 61        &  5    &  $<$0.02& $<$0.05 & w1, h1\\
NGC\,3627 & SAB(s)b;LINER Sy2 & 11h20m15.0s & +12d59m30s & +716         & 106       &  8    &  $<$0.21& $<$0.01 & w2, h6\\
NGC\,3628 & SAb pec sp;HII LINER & 11h20m17.0s & +13d35m23s & +825      & 103       &  7    &  $<$0.25&    -    & w2\\
\bf{Arp\,299}& merger            & 11h28m30.4s & +58d34m10s & +3073     & 111       &  5    &  126    &  13 & w2, w3, h3\\
\bf{NGC\,4038} & SB(s)m pec,merger  & 12h01m53.0s & -18d52m10s & +1634 & 76  &  28 &  8.2   & $<$0.07 & w4, h1\\
NGC\,4088 & SAB(rs)bc HII     & 12h05m34.2s & +50d32m21s & +746         & 52        &  16   &  $<$0.12&    -    & w2\\
NGC\,4102 & SAB(s)b?;HII LINER & 12h06m23.1s & +52d42m39s & +838        & 69        &  17   &  $<$0.16& $<$0.02 & w2, h2\\
NGC\,4254 & SA(s)c            & 12h18m49.6s & +14d24m59s & +2390        & 72        &  6    &  $<$1.81&   -     & w2\\
NGC\,4303 & SAB(rs)bc;HII Sy2 & 12h21m54.9s & +04d28m25s & +1586        & 62        &  13   &  $<$0.44&  -      & w2\\
NGC\,4321 & SAB(s)bc;LINER HII & 12h22m54.9s & +15d49m21s & +1581       & 58        &  12   &  $<$0.41& $<$0.02 & w2, h4\\
NGC\,4414 & SA(rs)c?;HII LINER & 12h26m27.1s & +31d13m25s & +711        & 68        &  17   &  $<$0.12& $<$0.007 & w2, h4\\
NGC\,4490 & SB(s)d pec        & 12h30m36.4s & +41d38m37s & +583         & 78        &  17   &  $<$0.08& $<$0.01 & w2, h2\\
NGC\,4501 & SA(rs)b;HII Sy2   & 12h31m59.2s & +14d25m14s & +2264        & 55        &  10   &  $<$0.69&    -    & w2\\
\bf{NGC\,4527}& SAB(s)bc;HII LINER & 12h34m08.5s & +02d39m14s & +1730   & 64        &  15   &  4      &    -    & w5\\
NGC\,4631 & SB(s)d            & 12h42m08.0s & +32d32m29s & +687         & 120       &  7    &  $<$0.17& $<$0.005 & w2, h4\\
NGC\,4666 & SABc;HII LINER    & 12h45m08.6s & -00d27m43s & +1524        & 77        &  10   &  $<$0.31&    -    & w2\\
NGC\,4736 & SA(r)ab;Sy2 LINER & 12h50m53.0s & +41d07m14s & +321      & 105          &  5    &  $<$0.03& $<$0.004 & w2, h2\\
NGC\,4826 & SA(rs)ab;HII Sy2  & 12h56m43.7s & +21d40m58s & +407       & 76          &  11   &  $<$0.03& $<$0.004 & w2, h6\\
\bf{NGC\,4945} & SB(s)cd: sp Sy2   & 13h05m27.5s & -49d28m06s & +547    & 620       &  -    &  165    & $<$0.03 & w6, h7\\
NGC\,5005 & SAB(rs)bc;Sy2 LINER & 13h10m56.2s & +37d03m33s & +952       & 59        &  13   &  $<$0.16&    -    & w2\\
NGC\,5055 & SA(rs)bc HII/LINER & 13h15m49.3s & +42d01m45s & +524        & 101       &  12   &  $<$0.05& $<$0.008 & w2, h2\\
NGC\,5128 (Cen A)& S0 pec Sy2        & 13h25m27.6s & -43d01m09s & +529  & 314       &  2    &  $<$0.02&   -     & w1\\
\bf{NGC\,5194} & SA(s)bc pec;HII Sy2.5 & 13h29m52.7s & +47d11m43s & +467   & 123    &  10   &  0.63   & $<$0.01 & w2, h2\\
NGC\,5236 & SAB(s)c;HII Sbrst & 13h37m00.9s & -29d51m56s & +494         & 213       &  34   &  $<$0.11& $<$0.01 & w2, h6\\
Arp\,220  & merger           & 15h34m57.1s & +23d30m11s & +5454        & 118       &  10   &  $<$4.01 &  580 & w2, h3, h9\\
NGC\,6000 & SB(s)bc:;HII Sbrst & 15h49m49.5s & -29d23m13s & +2178       & 59        &  25   &  $<$1.60& $<$0.06 & w2, h7\\
NGC\,6744 & SAB(r)bc LINER    & 19h09m46.1s & -63d51m27s & +857         & 86        &  15   &  $<$0.13&    -    & w1\\
NGC\,6946 & SAB(rs)cd;Sy2 HII & 20h34m52.3s & +60d09m14s & +51          & 128       &  11   &$<$0.0004& $<$0.00007& w2, h2\\
NGC\,7331 & SA(s)b LINER      & 22h37m04.1s & +34d24m56s & +806         & 82        &  14   &  $<$0.12& $<$0.02 & w2, h2\\
NGC\,7552 & SB(s)ab;HII LINER & 23h16m10.8s & -42d35m05s & +1628    & 102           &  13   &  $<$0.41& $<$0.05 & w1, h7\\
NGC\,7582 & SB(s)ab Sy2       & 23h18m23.5s & -42d22m14s & +1555       & 73         &  13   &  $<$0.40& $<$0.12 & w1, h1\\
NGC\,7793 & SA(s)d HII        & 23h57m49.8s & -32d35m28s & +247         & 56        &  7    & $<$0.002&    -    & w1\\
\\
\hline 
\end{tabular}
\end{center}
\scriptsize{$^a$ Source names with detected \hdueo \, masers are presented boldfaced.\\
$^b$ Classification, coordinates and LSR velocities are taken from the NASA/IPAC Extragalactic Database (NED).\\
$^{c}$ Flux densities at 100 $\rm{\mu m}$ are taken from the IRAS Point Source Catalog (Fullmer \& Lonsdale, \cite{fullmer89}).\\
$^{d}$ Isotropic luminosity thresholds are derived from $L/[\solum]= k \times S/[\rm{Jy}] \times \Delta \textit{v}/[\rm{km \ s^{-1}}] \times \textit{D}^{2}/[\rm{Mpc^{2}}]$, where $k$ is 0.023 and 0.0017 for \hdueo \, and OH masers respectively.\\
$^{e}$ References: (w1) present work; (w2) HPT; (w3) Tarchi et al. (\cite{tarchi07}); (w4) Darling et al. (\cite{darl08}); (w5) Braatz \& Gugliucci (\cite{braatz08}); (w6) Dos Santos \& L\`{e}pine (\cite{santos79}); (h1) Norris et al. (\cite{norris89}); (h2) Baan et al. (\cite{baan92}); (h3) Staveley-Smith et al. (\cite{stave87}); (h4) Schmelz \& Baan (\cite{schm88}); (h5) Gallimore et al. (\cite{galli96}); (h6) Unger et al. (\cite{unger86}); (h7) Staveley-Smith et al. (\cite{stave92}); (h8) Baan \& Irwin (\cite{baan95}); (h9) Baan et al. (\cite{baan82}).}
\label{maser}
\end{table*}

\section{Conclusions}
We have observed 12 sources out of a sample of 20 galaxies with $\rm{100 \, \mu m}$ IRAS point source flux densities $> 50 \rm{Jy}$ and declination $\rm{<-30^{\circ}}$ to search for water maser emission at 22\,GHz. The main results are:
\begin{enumerate}
\item By including the southern galaxies, our complete FIR-sample covers the entire sky.\\
\item Two new detections were obtained. One is a kilomaser with an isotropic luminosity, $L_{\rm{H_{2}O}}$, of $\rm{\sim5\, \solum}$ in the SB galaxy NGC\,3620. The other is a maser with $L_{\rm{H_{2}O}}\sim11 \,\solum$ in the merger system NGC\,3256.\\
\item Follow-up interferometric observations of both sources have been performed. In NGC\,3256, the maser is resolved into two spots offset from the two nuclei of the system hinting at an association with particularly vigorous star formation. For NGC\,3620, the maser emission coincides with the nuclear region. Also in this case an association with star formation activity is most likely, although an association with a Low Luminosity AGN cannot be a priori ruled out.\\
\item The high rate of water maser detections in the FIR sample encompassing northern and southern sources (23\%, 15/65) confirms a link between FIR flux density and maser phenomena as reported in HPT and CTH. Our results further reinforce the correlation between water masers and FIR luminosities for kilomasers while the correlation with megamasers is weaker. The latter is explained by the fact that megamasers in the nuclear region are not related to the overall star forming activity of their parent galaxy. Albeit this awaits a more extensive statistical dataset, an association between \hdueo \, and OH masers in the FIR sample is suggested for sources where at least one of the maser species has weak emission. However, there is no apparent correlation between \hdueo \, and OH megamasers. Given the relatively low number of sources, a systematic study of \hdueo \, megamaser sources in OH, and of OH megamasers in \hdueo \, would be highly desirable.\\
\end{enumerate}

\section*{Acknowledgments}
We wish to thank Shinji Horiuchi for performing part of the observations at Tidbinbilla and for useful discussions, and an anonymous referee for making useful suggestions. AT would like to thank Lincoln Greenhill and Paul Kondratko for helpful suggestions in the early stages of this project, and Alessandro Riggio for his advice on statistics. GS would like to thank Wouter Vlemmings for helpful comments. This research has made use of the NASA/IPAC Extragalactic Database (NED), which is operated by the Jet Propulsion Laboratory, Caltech, under contract with NASA. This research has also made use of NASA's Astrophysics Data System Abstract Service (ADS).The National Radio Astronomy Observatory is a facility of the National Science Foundation operated under cooperative agreement by Associated Universities, Inc. GS is a Member of the International Max Planck Research School (IMPRS) for Radio and Infrared Astronomy at the University of Bonn.\\


\begin{thebibliography}{}

\bibitem[1982]{baan82}
 Baan, W.A., Wood, P.A.D. \& Haschick, A.D., 1982, ApJ, 260, L49
\bibitem[1992]{baan92}
 Baan, W.A., Haschick, A. \& Henkel, C., 1992, AJ, 103, 728
\bibitem[1995]{baan95}
 Baan, W.A. \& Irwin, J.A., 1995, ApJ, 446, 602
\bibitem[1982]{batche82}
 Batchelor, R.A., Jauncey, D.L., Whiteoak, J.B., 1982, MNRAS, 200, 19
\bibitem[2009]{ben09}
 Bennert, N., Barvainis, R., Henkel, C., \& Antonucci, R., 2009, ApJ, astro-ph0901.0567
\bibitem[1997]{boker97}
 B\"{o}ker, T., Storey, J.W.V., Krabbe, A., \& Lehmann, T., 1997, PASP, 109, 827
\bibitem[1996]{braatz96}
 Braatz, J.A., Wilson, A.S., Henkel, C., 1996, ApJS, 106, 51
\bibitem[1997]{braatz97}
 Braatz, J.A., Wilson, A.S., Henkel, C., 1997, ApJS, 110, 321
\bibitem[2004]{braatz04}
 Braatz, J.A., Henkel, C., Greenhill, L.J. et al. 2004, ApJ, 617, 29
\bibitem[2007]{braatz07}
Braatz, J.; Kondratko, P.; Greenhill, L. et al. 2007, in IAU Symp., 242, 402
\bibitem[2008]{braatz08}
 Braatz, J.A. \& Gugliucci, N.E., 2008, ApJ, 678, 96
\bibitem[2008]{braatz008}
 Braatz, J.A., Reid, M.J., Greenhill, L.J. et al. 2008, ASPC, 395, 103
\bibitem[2005]{bru05}
 Brunthaler, A., Reid, M.J., Falcke, H. et al. 2005, Sci, 307, 1440
\bibitem[2006]{bru06}
 Brunthaler, A., Henkel, C., de Blok, W.J.G. et al. 2006, A\&A, 457, 109
\bibitem[2007]{bru07}
 Brunthaler, A., Reid, M.J., Falcke, H. et al. 2007, A\&A 462, 101
\bibitem[2008]{casta08}
 Castangia, P., Tarchi, A., Henkel, C., Menten, K.M., 2008, A\&A, 479, 111 (CTH)
\bibitem[1991]{caso91}
 Casoli, F., Dupraz, C., Combes, F., \& Kazes, I., 1991, A\&A, 251, 1
\bibitem[1990]{cesa90}
 Cesaroni, R. 1990, A\&A, 233, 513
\bibitem[2007]{darl07}
Darling, J. 2007, in IAU Symp., 242, 417
\bibitem[2008]{darl08}
 Darling, J., Brogan, C., \& Johnson, K., 2008, ApJ, 685, 39
\bibitem[1961]{vaun61}
 de Vaucouleurs, G. \& de Vaucouleurs, A. 1961, Mem. R. Astron. Soc., 68, 69
\bibitem[1979]{santos79}
 Dos Santos, P.M., \&  L\`{e}pine, J.R.D., 1979, Nature, 278, 34
\bibitem[1994]{doyon94}
 Doyon, R., Joseph, R.D., \& Wright, G.S. 1994, ApJ, 421, 101
\bibitem[1996]{elfhag96}
 Elfhag, T., Booth, R.S., H\"{o}glund, B. et al. 1996, A\&AS, 115, 439
\bibitem[2003]{eng03}
 English, J., Norris, R.P., Freeman, K.C. \& Booth, R.S. 2003, ApJ, 125, 1134
\bibitem[1989]{for89}
 Forster, J.R. \& Caswell, J.L., 1989, A\&A, 213, 339
\bibitem[1989]{fullmer89}
 Fullmer, L., \& Lonsdale, C. 1989, Cataloged Galaxies and Quasars Observed in the IRAS Survey, Version 2, JPL D-1932
\bibitem[1996]{galli96}
 Gallimore, J.F., Baum, S.A., O'Dea, C.P. et al. 1996, ApJ, 462, 740
\bibitem[1997]{galli97}
 Gallimore, J.F., Baum, S.A., O'Dea, C.P. \& Claussen, M. 1997, IAUJD, 21E, 10
\bibitem[1997]{green97}
 Greenhill, L.J., Moran, J.M., \& Herrnstein, J.R., 1997, ApJ, 560, L37
\bibitem[2003]{green03}
 Greenhill, L.J., Kondratko, P.T., Lovell, J.E.J. et. al. 2003, ApJ, 582L,11 
\bibitem[2001]{hagiwara01}
 Hagiwara, Y., Henkel, C., Menten, K.M., Nakai, N. 2001, ApJ, 560, L37
\bibitem[2003]{hagiwara03}
 Hagiwara, Y., Diamond, P.J., Miyoshi, M., et al. 2003, MNRAS, 344, L53
\bibitem[1986]{henkel86}
 Henkel, C., Wouterloot, J.G.A. \& Bally, J. 1986, A\&A, 155, 193
\bibitem[1990]{hen90}
Henkel, C. \& Wilson, T.L. 1990, A\&A, 229,431
\bibitem[2005a]{henkel05a}
 Henkel, C., Braatz, J.A., Tarchi, A. et al. 2005a, ApSS, 295, 107
\bibitem[2005b]{henkel05b} 
 Henkel, C., Peck, A. B., Tarchi, A. et al. 2005b, A\&A, 436, 75 (HPT)
\bibitem[1999]{her99}
 Herrnstein, J. R., Moran, J.M., Greenhill, L.J. et al. 1999, Nature, 400, 539
\bibitem[1987]{Ho87}
 Ho, P.T.P., Martin, R.N., Henkel, C. et al. 1987, ApJ, 320, 663
\bibitem[1974]{hogbom74}
 H\"{o}gbom, J.A., 1974, A\&AS, 15, 417
\bibitem[2008]{imp08}
 Impellizzeri, C.M.V., Kean, J.P., Castangia, P. et al 2008, Nature, 456, 927
\bibitem[1981]{jaf81}
 Jaffe, D.T., Guesten, R., \& Downes, D. 1981, ApJ, 250, 621
\bibitem[2004]{jen04}
 Jenkins, L.P., Roberts, T.P., Ward, M.J.,\& Zezas, A. 2004, MNRAS, 352, 1335
\bibitem[2000]{lipari00}
 L\'{\i}pari, S., D\'{\i}az, R., Taniguchi, Y. et al. 2000, AJ, 120, 645
\bibitem[2002]{lira02}
 Lira, P., Ward, M., Zezas, A. et al. 2002, MNRAS, 330, 259
\bibitem[2008]{lira08}
 Lira, P., Gonzalez-Corvalan, V., Ward, M., \& Hoyer, S., 2008, MNRAS, 384, 316 
\bibitem[2005]{lo05}
 Lo, K.-Y. 2005, ARA\&A, 43, 625
\bibitem[2006]{kondra06}
 Kondratko, P.T., Greenhill, L.J., Moran, J.M. et al. 2006, ApJ, 638, 100
\bibitem[1996]{Kont96}
 Kotilainen, J.K., Moorwood, A.F.M., Ward, M.J., \& Forbes, D.A. 1996, A\&A, 305, 107
\bibitem[2009]{min09}
 Minter, T., ``The Proposer's Guide for the Green Bank Telescope'', 2009
\bibitem[1995]{miy95}
 Miyoshi, M., Moran, J., Herrnstein, J.R. et al. 1995, Nature, 373, 127
\bibitem[2003]{neff03}
 Neff, S.G., Ulvestad, J.S., \& Campion, S.D. 2003, ApJ, 599, 1043
\bibitem[1989]{norris89}
 Norris, R.P., Gardner, F.F., Whiteoak, J.B. et al. 1989, MNRAS, 237, 673
\bibitem[1995]{norris95}
 Norris, R.P. \& Forbes, D.A. 1995, ApJ, 446, 594 (NF95)
\bibitem[2006]{per06}
Perley, R., Hayward, B., Butler, B. et al. 2006, EVLA memo 103; www.aoc.nrao.edu/evla/geninfo/memoseries
\bibitem[1998]{phill98}
 Phillips, C.J., Norris, R.P., Ellingsen, S.P. \& Rayner, D.P. 1998, MNRAS, 294, 265
\bibitem[2009]{reid09}
 Reid, M.J., Braatz, J.A., Condon, J.J. et al. 2009, [arXiv:astro-ph/0811.4345]
\bibitem[2006]{saka06}
 Sakamoto, K., Ho, P.T.P., Peck, A.B. et al. 2006, ApJ, 644, 862
\bibitem[2003]{sand03}
 Sanders, D.B., Mazzarella, J.M., Kim, D.-C. et al. 2003, ApJ, 126, 1607
\bibitem[1989]{sar89}
 Sargent, A.I., Sanders, D.B., \& Phillips, T.G. 1989, ApJ, 346, L9
\bibitem[1988]{schm88}
 Schmelz, J.T. \& Baan, W.A. 1988, AJ, 95, 672
\bibitem[1978]{schw78}
 Schwartz, D., 1978, PASP, 90, 393
\bibitem[1987]{stave87}
 Staveley-Smith, L., Cohen, R.J., Chapman, J.M. et al. 1987, MNRAS, 226, 689
\bibitem[1992]{stave92}
 Staveley-Smith, L., Norris, R.P., Chapman, J.M. et al. 1992, MNRAS, 258, 725
\bibitem[2002]{tarchi02}
 Tarchi, A., Henkel, C., Peck, A.B., et al. 2002, A\&A, 389, 39
\bibitem[2007]{tarchi07}
 Tarchi, A., Castangia, P., Henkel, C. \& Menten, K.M. 2007, NewAR, 51, 67
\bibitem[1998]{tro98}
 Trotter, A.S., Greenhill, L.J., Moran, J.M. et al. 1998, ApJ, 495, 740
\bibitem[1986]{unger86}
 Unger, S.W., Chapman, J.M., Cohen, R.J. et al. 1986, MNRAS, 220, 1

\end{thebibliography}
\end{document}